\documentclass[12pt, a4paper]{article}
\pdfoutput=1
\usepackage[dvipdfmx]{graphicx}
\usepackage{amssymb}
\usepackage{amsmath}
\usepackage{bm}
\usepackage{color}
\usepackage{cite}
\usepackage{epstopdf}            
\usepackage{epsfig}
            
\newcommand{\beq}{\begin{equation}}
\newcommand{\eeq}{\end{equation}}

\usepackage[colorlinks=true, linkcolor=black, citecolor=black,
urlcolor=black]{hyperref}

\begin{document}

\begin{titlepage}

\begin{center}

{\Large
{\bf 
Hidden sector behind the CKM matrix
}
}

\vskip 2cm

Shohei Okawa$^1$
and
Yuji Omura$^{2}$

\vskip 0.5cm

{\it $^1$
Department of Physics, Nagoya University, Nagoya 464-8602, Japan}\\[3pt]
{\it $^2$
Kobayashi-Maskawa Institute for the Origin of Particles and the
Universe, \\ Nagoya University, Nagoya 464-8602, Japan}\\[3pt]

\vskip 1.5cm

\begin{abstract}
The small quark mixing, described by the Cabibbo-Kobayashi-Maskawa (CKM) matrix in the Standard Model,
may be a clue to reveal new physics around the TeV scale.
We consider a simple scenario that extra particles in a hidden sector radiatively
mediate the flavor violation to the quark sector around the TeV scale and effectively realize the observed CKM matrix. 
The lightest particle in the hidden sector, whose contribution to
the CKM matrix is expected to be dominant, is a good dark matter (DM) candidate.
There are many possible setups to describe this scenario, so that we investigate 
some universal predictions of this kind of model, focusing on the contribution of DM to the quark mixing and flavor physics. In this scenario, there is an explicit relation between the CKM matrix 
and flavor violating couplings, such as four-quark couplings, because both are radiatively induced by the particles in the hidden sector. Then, we can explicitly find the DM mass region and the size of Yukawa couplings between the DM and quarks, based on the study of flavor physics and DM physics. In conclusion, we show that DM mass in our scenario is around the TeV scale, and the Yukawa couplings are between ${\cal O}(0.01)$ and ${\cal O}(1)$. 
The spin-independent DM scattering cross section is estimated as ${\cal O}(10^{-9})$ [pb]. An extra colored particle is also predicted at the ${\cal O}(10)$ TeV scale.
\end{abstract}

\end{center}
\end{titlepage}

\section{Introduction}
The flavor structure of the Standard Model (SM) is one of mysteries, which are
expected to be solved by extending the SM. In the SM, there are three generations
in both quark and lepton sectors, and the difference among the generations is the size of the fermion masses.
The fermion masses are dynamically generated by the spontaneous electroweak (EW) symmetry breaking,
and the observed masses and mixing are given by the Yukawa couplings with the Higgs field in the SM.
We know that the Yukawa couplings have to realize the large mass hierarchies and
the small quark mixing. This unique form of the Yukawa matrix may be a clue to reveal the new physics above the EW scale.

If the Yukawa couplings are ignored in the SM Lagrangian, the flavor symmetry to rotate generations and phases of quarks is restored. Of these, the rotation symmetry of the generations is broken by quark mass terms;
on the other hand, the symmetry to rotate the quark phases is respected even in the mass terms.
The phase rotation is explicitly broken only by the Cabibbo-Kobayashi-Maskawa (CKM) matrix in the weak interaction. According to the experimental results, the CKM matrix is close to the $3 \times 3$ identity matrix but has small mixing angles. These small mixing angles may imply that the flavor symmetry, especially to rotate the quark phases,
is respected at high energy. If this is the case, new physics exists above the EW scale in order to
break the flavor symmetry spontaneously and generate the realistic quark mixing.
This simple scenario, however, possibly suffers from constraints from flavor violating processes.
If the flavor symmetry breaking of the Yukawa couplings is generated at the tree level,
large flavor changing neutral currents (FCNCs) are generally induced and the model is easily excluded.
Therefore, we need consider a scenario that some new particles mediate the flavor symmetry breaking to the quark sector
at a loop level.

We have another strong motivation to desire new physics above the EW scale, that is, the results of the cosmological observations proposed by the WMAP and Planck collaborations \cite{DMexperiment2,DMexperiment}.
They suggest that dark energy and dark matter (DM) dominate our universe and
the amount of DM is about five times bigger than the visible particles.
There are a lot possibilities for DM and one possible DM candidate is
a Weakly-Interacting Massive Particle (WIMP) which resides around TeV scale.
Then, we can expect that there is a direct connection between DM and the origin of the quark matrix.

\begin{figure}[!t]
\begin{center}
{\epsfig{figure=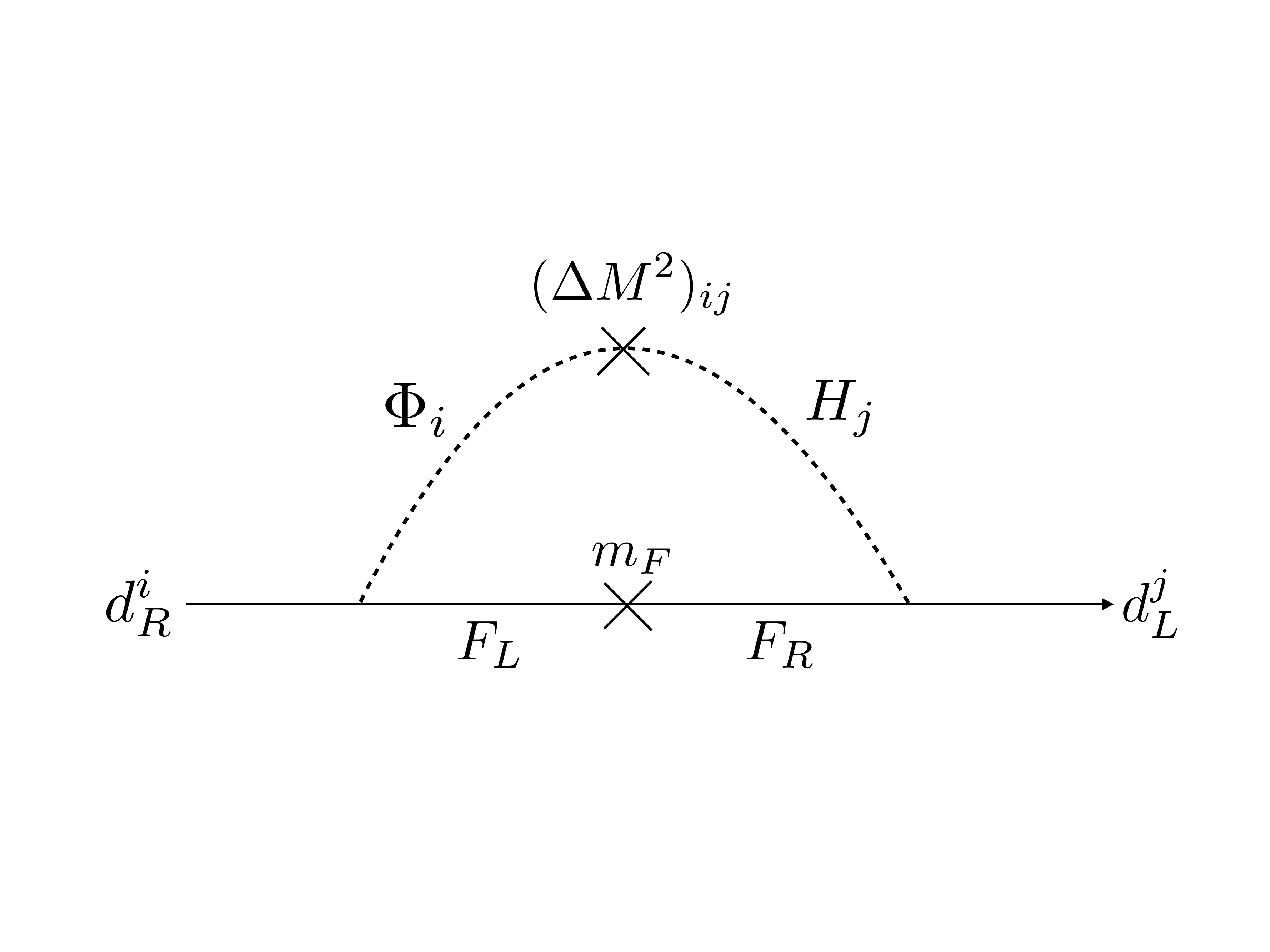,width=0.5\textwidth}}
\caption{Rough sketch of our idea to generate the flavor violation. $\Delta M_{ij}$ breaks the flavor symmetry.}
\label{fig;idea}
\end{center}
\end{figure}

Motivated by those mysteries, in this paper, we consider a simple scenario that extra particles, including DM, radiatively
mediate the flavor symmetry breaking to the quark sector around the TeV scale and effectively realize the observed quark mixing. In our scenario, the flavor symmetry to rotate the quark phases in each generation is conserved at high energy in both of the up-type and down-type quark sectors. At the ${\cal O}(10)$ TeV scale, the flavor symmetry breaks down in a hidden sector.
We also introduce an extra heavy quark and scalars charged under the flavor symmetry to mediate the flavor symmetry breaking to the quark sector.
The mediators have flavor symmetric Yukawa couplings with only down-type quarks, but the fields to break
the flavor symmetry do not couple to any SM fermions directly.
The scalar mediators are only the fields to couple with the symmetry breaking fields, 
and then radiatively mediate the flavor symmetry breaking to the quark sector.
We do not construct any explicit model for the flavor symmetry breaking, assuming that the effect of the symmetry breaking appears in the mass matrix
of the scalar mediators. The rough sketch of this idea is shown in Fig. \ref{fig;idea}. 
$\Phi_i$, $H_i$ and $F$ correspond to the SM-singlet, the EW charged scalars, and the extra heavy quark as the mediators. $(\Delta M)_{ij}$ is the part of the mass matrix for the scalars and denotes the flavor symmetry breaking effect.
Note that we can find many similar setups motivated by the origin of the quark mixing\cite{Barbieri:1980vc,Barbieri:1980tz,Barbieri:1981yw,Barbieri:1981yy,Kramer:1981sq,Balakrishna:1988ks,Balakrishna:1987qd,Lahanas:1982et,Masiero:1983ph,Barr:1984pk,Kagan:1989fp,Baumgart:2014jya,Altmannshofer:2014qha,Borzumati:1999sp,Ferrandis:2004ng,Ferrandis:2004ri,Crivellin:2010ty,Crivellin:2011sj,Thalapillil:2014kya,He:1989er,Ma:2013mga, Ma:2014yka,Natale:2016xob,Nomura:2016emz,Kownacki:2016hpm,CarcamoHernandez:2016pdu}. 
In the present work, we consider a simple setup motivated by DM as well as the explanation of the quark mixing, 
and survey predictions of this kind of model.
The results could be applied to many concrete models that radiatively induce the quark mixing.

The lightest neutral particle among the scalars, $\Phi_i$ and $H_i$,
is a good DM candidate. The one-loop contribution to the
down-type quark Yukawa couplings in Fig. \ref{fig;idea} is probably dominated by the
contribution of the diagram involving the DM, because of the relatively light mass. 
Then, we simply focus on the physics of the DM and estimate the 
size of the predicted quark mixing and quark masses. As mentioned above, there are 
many possible setups to describe this kind of scenario \cite{Barbieri:1980vc,Barbieri:1980tz,Barbieri:1981yw,Barbieri:1981yy,Kramer:1981sq,Balakrishna:1988ks,Balakrishna:1987qd,Lahanas:1982et,Masiero:1983ph,Barr:1984pk,Kagan:1989fp,Baumgart:2014jya,Altmannshofer:2014qha,Borzumati:1999sp,Ferrandis:2004ng,Ferrandis:2004ri,Crivellin:2010ty,Crivellin:2011sj,Thalapillil:2014kya,He:1989er,Ma:2013mga, Ma:2014yka,Natale:2016xob,Nomura:2016emz,Kownacki:2016hpm,CarcamoHernandez:2016pdu}. 
We could find some universal predictions, according to this simple assumption that
the observed CKM matrix is originated from the one-loop corrections involving the DM.

Interestingly, the one-loop correction is roughly estimated as ${\cal O}(10^{-3})$ when the
coupling between DM and quarks is a little smaller than ${\cal O}(1)$. 
It is close to the order of the strange quark mass divided by the vacuum expectation value (VEV)
of the Higgs field. In order to realize the CKM matrix, the required correction to the down-type Yukawa matrix
is also between ${\cal O}(10^{-6})$ and ${\cal O}(10^{-4})$, so that the couplings between DM and quarks should be in the range between ${\cal O}(0.01)$ and ${\cal O}(1)$, depending on the DM mass.
Then, we need not worry about the triviality bound concerned with the divergence of couplings
and we can predict a sizable interaction between DM and nuclei.

 %-----------------
%------------------------------
 
Another important prediction is a direct connection between the CKM matrix and flavor violating couplings, such as four-quark couplings.
In this kind of setup, the CKM matrix and the flavor-violating couplings have the same source,
so that sizable deviations from the SM values in the flavor violating processes are predicted.
It is known that the $\Delta F=2$ processes and the electric dipole moment (EDM)
give stringent bounds to new physics contributions, so that
the flavor physics constrains the mass scale of the heavy quark and the extra scalars.
In our work, we see that the non-vanishing CP phases in the Wilson coefficients 
of the $\Delta S=2$ and the EDM operators are unavoidable, 
and then we conclude that the extra quark mass should be not
less than about 10 TeV. Taking into account the vacuum stability and the relic density of DM, 
the DM mass range is predicted to be between about 1 TeV and 10 TeV. 
We also find an explicit prediction of the spin-independent DM scattering cross section in Sec. \ref{sec4}:
$\sigma_{\rm SI} \simeq 1.7 \times 10^{-9}$ [pb]. Then, we conclude that our DM candidate 
can be tested by the future prospect of the XENON1T experiment \cite{XENON1T}.

In Sec. \ref{sec2}, we introduce our setup and explain the underlying theory of our scenario.
Then, we discuss how to realize the observed Yukawa couplings in the SM.
Since there are correlations between the predicted CKM matrix and the contributions to the flavor violating processes
involving DM, we can explicitly derive the DM mass region. This study is given in Sec. \ref{sec3}. Based on the study in Sec. \ref{sec3}, we discuss
the DM physics in Sec. \ref{sec4}. In the last section, we summarize our results 
and give a short comment on the other setup of the mediation and hidden sectors, motivated by the origin of
the quark mass matrices.

\section{Setup}
\label{sec2}
We propose a scenario that the small quark mixing in the SM
is originated from flavor symmetry breaking in a hidden sector.
In our assumption, there is a flavor symmetry to rotate quark phases in each generation.
The flavor symmetry is spontaneously broken by some fields in the hidden sector at some scale. 
The SM quarks do not directly couple with any fields to break the flavor symmetry, but 
there exist an extra quark and extra scalars to mediate the breaking effect to the quark sector. 
The fields to mediate the breaking effect do not contribute to the dynamics of the flavor symmetry breaking,
but the masses and the mass eigenstates are affected by the symmetry breaking.
%%%%%%%%%%%%%%%%%%%%%%%%%%%%%%%%%%%%%%%%%%%%%%%%%%%%%%%%%%%
\begin{table}[!t]
\begin{center}
\begin{tabular}{cccccc}
\hline
\hline
Fields & ~~spin~~   & ~~$\text{SU}(3)_c$~~ & ~~$\text{SU}(2)_L$~~  & ~~$\text{U}(1)_Y$~~ &~~flavor charge~~     \\ \hline
  $\Hat Q^{i}_L$  &1/2 & ${\bf 3}$         &${\bf2}$      &       $1/6$   & $q_i$         \\  
   $\Hat u^{i}_R$  &1/2 & ${\bf 3}$         &${\bf1}$      &       $2/3$   & $q_i$         \\  
   $\Hat d^{i}_R$  & 1/2 & ${\bf 3}$        &${\bf1}$      &         $-1/3$      & $q_i$          \\ 
     $ H$  & 0 & ${\bf1}$         &${\bf 2}$      &            $1/2$   & $0$         \\  \hline   
%        $\Hat F$  & 1/2 & ${\bf 3}$        &${\bf1}$      &         $-1/3$      & $q_F$          \\ \hline
%    $ \Phi_{1,2}$  & 0 & ${\bf1}$         &${\bf 1}$      &            $0$   & $q^\Phi_{1,2}$         \\
 %  $ X$ (DM) & 0 & ${\bf1}$         &${\bf 1}$      &            $0$   & $q_X$         \\ \hline \hline
\end{tabular}
\end{center}
\label{table1}
\caption{SM particles with flavor charges. }
\end{table} 
%%%%%%%%%%%%%%%%%%%%%%%%%%%%%%%%%%%

First, let us summarize the matter content of the quark sector. 
The fields in the flavor base are shown in Table \ref{table1}.
$\Hat Q^{i}_L$, $\Hat u^{i}_R$, and $\Hat d^{i}_R$  ($i=1, \,2, \,3$) are the left-handed quarks, right-handed up-type and
down-type quarks in the flavor base. 
We introduce a flavor symmetry to rotate the quark phases and assign a flavor charge ($q_i$) to each quark
such that the flavor symmetry is conserved even in the Yukawa interaction with the Higgs doublet ($H$):
\begin{equation}
\label{eq;Yukawa}
 V_{{\rm Y}}=y^u_{i} \overline{\Hat Q^i_L} \widetilde{H} \Hat u^i_R + y^d_{i} \overline{\Hat Q^i_L} H \Hat d^i_R. 
\end{equation}
The Yukawa couplings, $y^u_{i}$ and $y^d_{i}$, are in the diagonal forms, so that 
no quark flavor mixing appears at this level.
Note that the flavor symmetry is not needed to be continuous symmetry, like U(1).
It is not specified in our paper, assuming that the flavor symmetry is broken only in the hidden sector above the EW scale.  

In the hidden sector there are extra fields to break the flavor symmetry and mediate the breaking effect.
Let us introduce flavor-charged scalars, $\Phi_i$ and $H_i$, together with a flavor-singlet colored particle, $F$.
The SM charges of $F$ are the same as the ones of right-handed down-type quarks. 
$\Phi_i$ and $H_i$ are SM-singlet complex scalar and SU(2)$_L$ doublets charged under the flavor symmetry, respectively.
Then, we write down the flavor conserving Yukawa couplings between the extra fields and down-type quarks:
\begin{equation}
\label{eq;Yukawa-extra}
 V_{{\rm extra}}=\Hat \lambda_{i} \overline{F_L} \Phi^\dagger_i \Hat d^i_R + \Hat \kappa_{i} \overline{\Hat Q^i_L} H_i  F_R. 
\end{equation}

Here, we simply assume that the flavor symmetry is broken by some fields in the hidden sector except for $\Phi_i$ and
$H_i$, and the mass eigenstates of $\Phi_i$ and
$H_i$ are fixed by the scalar potential involving the fields to break the symmetry.\footnote{The flavor symmetry might be explicitly broken. We do not specify the structure of the hidden sector. In the case that the flavor symmetry is spontaneously broken in the hidden sector, we can easily construct a model introducing extra flavored SM-singlet fields, $\varphi_{i}$. The potential for the flavor symmetry breaking is, for instance, given by $V_{SB}=- \mu^2_i |\varphi_{i}|^2 +  \lambda_i  |\varphi_{i}|^4$, and each $\varphi_{i}$ develops non-vanishing VEV. Note that $\varphi_{i}$ could be real scalars, depending on the flavor symmetry. } 
Then, we rewrite the scalars in the flavor base with the mass eigenstates denoted by $X$, $\phi_{1,2}$, $H_D$, and $H^\prime_{1,2}$:
\beq
\Phi_i = c^X_i X + c^{\phi_1}_i \phi_1+ c^{\phi_2}_i \phi_2, \, H_i = c^D_i H_D + c^{H^\prime_1}_i H^\prime_1+ c^{H^\prime_2}_i H^\prime_2.
\eeq
Each of the coefficients is given by the mass matrix including $(\Delta M)_{ij}$ in Fig. \ref{fig;idea}.
All of the scalars radiatively contribute to the quark mixing through the Yukawa coupling in Eq. (\ref{eq;Yukawa-extra}).
The size of the contribution from each scalar would depend on the detail of models.
In fact, we can consider many setups to realize the observed quark mass matrix radiatively in the
framework of the Grand Unified Theory \cite{Barbieri:1980vc,Barbieri:1980tz,Barbieri:1981yw,Barbieri:1981yy},
Left-Right symmetric models \cite{Kramer:1981sq,Balakrishna:1988ks,Balakrishna:1987qd}, supersymmetric models
\cite{Lahanas:1982et,Masiero:1983ph,Barr:1984pk,Kagan:1989fp,Baumgart:2014jya,Altmannshofer:2014qha,Borzumati:1999sp,Ferrandis:2004ng,Ferrandis:2004ri,Crivellin:2010ty,Crivellin:2011sj,Thalapillil:2014kya}, and
flavor symmetric models \cite{He:1989er,Ma:2013mga, Ma:2014yka,Natale:2016xob,Nomura:2016emz,Kownacki:2016hpm,CarcamoHernandez:2016pdu}. 
Our main motivations are, however, to find the connection between DM and the quark mixing in the SM
and to look for universal predictions of this kind of model.
Therefore, we especially concentrate on the case that the light scalars dominantly contribute to the quark mixing,
and the lightest scalar is a DM candidate.
In particular, we focus on a minimal setup to realize the observed quark mass matrix; that is, there are only two kinds of light scalars, $X$ and $H_D$, in our simplified model.
Assuming that $X$ and $H_D$ are relatively lighter than the others, we can approximately simplify the Yukawa couplings as,
\begin{equation}
\label{eq;Yukawa-extra-2}
 V^{\rm app}_{{\rm extra}}= \lambda_{i} \overline{F_L} X^\dagger \Hat d^i_R + \kappa_{i} \overline{\Hat Q^i_L} H_D  F_R +etc.. 
\end{equation}
We discuss the physics in our scenario, using these Yukawa couplings only: $\lambda_{i}$ and $\kappa_{i}$.
We will give some comments on the contributions of $\phi_{1,2}$ and $H^\prime_{1,2}$.
The charge assignment of the main fields for the mediation is summarized in Table \ref{table2}.
Note that dark charges are also assigned to $F$, $H_D$ and $X$, to distinguish them from the SM particles.
Thanks to the dark charge, $X$ and/or the neutral component of $H_D$ can be stable and good dark matter candidates to dominate our universe.

\begin{table}[thb]
\begin{center}
\begin{tabular}{cccccc}
\hline
\hline
Fields & ~~spin~~   & ~~$\text{SU}(3)_c$~~ & ~~$\text{SU}(2)_L$~~  & ~~$\text{U}(1)_Y$~~ & Dark charge    \\ \hline   
  $F$  & 1/2 & ${\bf 3}$        &${\bf1}$      &         $-1/3$        &   $+1$     \\ \hline
   $H_D$ & 0 & ${\bf1}$         &${\bf 2}$      &            $1/2$      &   $-1$      \\
   $X$ & 0 & ${\bf1}$         &${\bf 1}$      &            $0$      &   $-1$      \\ \hline \hline
\end{tabular}
\end{center}
\label{table2}
\caption{Extra particles in the mediation sector. }
\end{table} 
%%%%%%%%%%%%%%%%%%%%%%%%%%%%%%%%%%%%%%%%%%%%
%%%%%%%%%%%%%%%%%%%%%%%%%%%%%%%%%%%%%%%%%%%%

The scalar fields couple with the SM Higgs field as well, and 
the couplings, in addition to $V^{\rm app}_{{\rm extra}}$, are
\begin{eqnarray}
 \Delta V^{\rm app}_{{\rm extra}}&=& m_{F} \overline{F} F + A \, X H^\dagger_D H + m^2_{X} |X|^2 + m^2_{H} |H_D|^2.
%&+& \sum^2_{a=1}\lambda^{a}_{i} \overline{\Hat F_L} \Phi_a \Hat d^i_R+\sum^2_{a=1}\Hat \lambda^{a}_{i} \overline{\Hat F_L} \Phi^\dagger_a \Hat d^i_R+\lambda^{X}_{i} \overline{\Hat F_L} X \Hat d^i_R+\widetilde \lambda^{X}_{i} \overline{\Hat F_L} X^\dagger \Hat d^i_R+h.c..
\label{Yukawa1}
\end{eqnarray}
Now, we expect that $X$ and $H_D$ do not develop non-vanishing VEVs because of the positive $m^2_{X}$ and $m^2_{H}$. $A$ is a trilinear coupling, that is effectively induced after the flavor symmetry breaking.
The $A$ term couples the SM Higgs and the mediators, so that
plays an important role in generating the CKM matrix.

Note that the mass terms of $X$ and $H_D$ have a lower bound from the condition for the vacuum stability.
If the trilinear coupling, $A$, is too large compared to $m_X$ and $m_{H}$,
unstable directions would appear at the origin.
Then, we find the following condition for the mass terms:
\beq
\label{vacuum}
m^2_X m^2_{H} > \frac{1}{2} A^2 v^2,
\eeq 
where $v$ denotes the VEV of the Higgs field: $\langle H^T \rangle=(0, \, v/\sqrt{2}) $.

%%%%%%%%%%%%%%%%%%%%%%%%%%%%%%
\subsection{Realization of the realistic Yukawa couplings}
\begin{figure}[!t]
\begin{center}
{\epsfig{figure=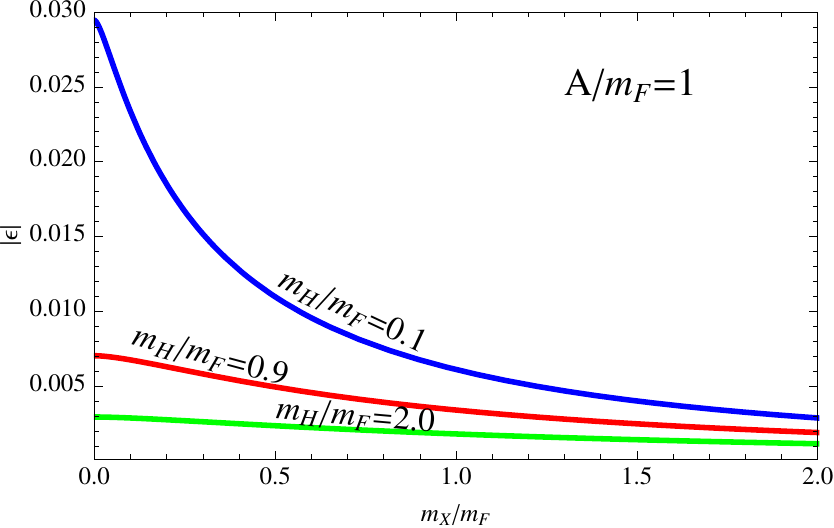,width=0.5\textwidth}}
\caption{$m_X/m_F$ vs. $|\epsilon|$ with $A/m_F=1$ and $m_H/m_F=0.1$ (blue), $0.9$ (red), and $2.0$ (green), respectively.  }
\label{fig;scenario1}
\end{center}
\end{figure}

%%%%%%%%%%%%%%%%%%%%%%%%%%%%%%
We discuss how the quark mixing is generated in our scenario.
The Yukawa couplings at the tree-level are given by Eq. (\ref{eq;Yukawa}), so that
they lead the diagonal mass matrices for up-type and down-type quarks after the EW symmetry breaking:
\beq
\label{eq;mass matrix0}
(M_u)_{ij} = \frac{y^u_i}{\sqrt 2} v \, \delta_{ij}, \, (M^{(0)}_d)_{ij} = \frac{y^d_i}{\sqrt 2} v \, \delta_{ij}.
\eeq
In our scenario, the flavor symmetry is broken in the hidden sector, and $F$, $X$ and $H_D$ decouple with the quark sector above the EW scale.
Then, the small quark mixing is effectively generated via the Yukawa couplings in Eq. (\ref{eq;Yukawa-extra-2}).
According to the one-loop correction as shown in Fig. \ref{fig;idea}, we obtain the
mass matrix for the down-type quarks in the form of
\beq
\label{eq;mass matrix}
(M_d)_{ij} = (M^{(0)}_d)_{ij}+ \frac{v}{\sqrt 2} \,  \epsilon  \, \kappa_{i} \lambda_j,
\eeq
where $\epsilon$ is the factor that comes from the one-loop correction:
\beq
\epsilon = \frac{1 }{16 \pi^2} \frac{A}{m_F} \, {\cal Y} (m^2_H/m^2_F, m^2_X/m^2_F).
\eeq
${\cal Y}(x, \, y)$ is given by 
\beq
 {\cal Y}(x, \, y)=\frac{x(y-1) \ln x -y(x-1) \ln y}{(x-1)(x-y)(y-1)}.
\eeq
$m_F$, $m_X$ and $m_H$ are the masses of the fields, $F$, $X$ and $H_D$, respectively.
$A$, $m_X$, and $m_H$ are expected to be around the flavor symmetry breaking scale.
The origin of $m_F$ may be independent of the breaking scale. 
$|\epsilon|$ could be estimated as ${\cal O} (10^{-3})$ when $A/m_F$, $m_X/m_F$ and $m_H/m_F$ are larger than ${\cal O}(0.1)$.
Fig. \ref{fig;scenario1} shows $m_X/m_F$ vs. $|\epsilon|$, assuming $m_H/m_F=0.1$ (blue), $0.9$ (red), $2.0$ (green) and  $A/m_F=1$. We see that $|\epsilon|$ is expected to be between ${\cal O} (10^{-3})$ and ${\cal O} (10^{-2})$ in the parameter region. 
%-----------------
%------------------------------
The required size of Yukawa couplings for the down-type quarks are less than ${\cal O} (10^{-2})$, so that
the size of the loop correction can be compatible with the required values for the down-type quark masses, as well as the CKM matrix.
Note that $\kappa_{i}$ and $\lambda_i$ are complex parameters, in principle, so CP phases are also generated by this dynamics.

Now, we define the mass eigenstates and derive the relation between the realistic mass matrix and the extra Yukawa couplings: $\kappa_{i}$ and $\lambda_i$.
The Yukawa matrix for the up-type quarks is in the diagonal form in Eq. (\ref{eq;mass matrix0}). Precisely speaking, 
there would be additional contributions from the wave function renormalization factor and
the loop correction involving $\phi_{1,2}$ and $H^\prime_{1,2}$.
The former is suppressed by $y^d_i$ in the mass matrix and the later is assumed to be sub-dominant in our scenario.\footnote{Fig. \ref{fig;scenario1} shows that we could obtain at least 10 \% suppressions compared to the contributions of $X$ and $H_D$, if the masses of $\phi_{1,2}$ and $H^\prime_{1,2}$ are larger than $m_F$. }
Then, we approximately derive the following relation, using the mass matrix in Eq. (\ref{eq;mass matrix}):
\beq
\label{eq;relation}
(m_d)_{i} \, \delta_{ij}= \frac{v}{\sqrt 2}  \left \{ \left ( V y^d V^\dagger_R \right )_{ij}  +   \epsilon  \, \left (V \kappa \right )_i \left ( \lambda V^\dagger_R  \right )_j  \right  \}.
\eeq
$(m_d)_i$ denote the quark masses: $(m^d_1,\, m^d_2, \, m^d_3)=(m_d,\, m_s, \, m_b)$.
$V$ is the CKM matrix and $V_R$ is the diagonalizing matrix which rotates right-handed down-type quarks.
Here, $V \kappa$ and $\lambda V^\dagger_R$ are the three-dimensional vectors, and they correspond
to the Yukawa couplings with DM (scalars) and $F$ in the mass base: 
\beq
\label{eq;Yukawa_extra}
V^{\rm ex}_{\rm Y}= \left ( \lambda V^\dagger_R  \right )_i \overline{F_L} X^\dagger d^i_R + \left (V \kappa \right )_i \overline{ d^i_L} H^0_D  F_R,
\eeq
where $ d^i_L$ and $ d^i_R$ are the mass eigenstates. We define $\widetilde \kappa_i \equiv V_{ij} \kappa_j$ and $\widetilde \lambda_i =  \lambda_j V^\dagger_{R \, ji}  $. In our notation, the mass eigenstates are $(d^1, \, d^2, \, d^3)=(d, \, s, \, b)$.
As we see in Sec. \ref{sec3}, the flavor violating couplings that contribute to the $\Delta F=2$ processes are also generated by the Yukawa couplings via the box diagram involving $F$, $X$ and $H_D$. 
According to the relation in Eq. (\ref{eq;relation}), $\widetilde \kappa_i$ and $\widetilde \lambda_i$ are explicitly related to
the CKM matrix and the quark masses, so that we can expect to obtain explicit predictions to the flavor violating processes.
Before the detailed analyses of the flavor physics, let us discuss the consistency with the realistic Yukawa couplings
and estimate the size of $\widetilde \kappa_i$ and $\widetilde \lambda_i$ lead by Eq. (\ref{eq;relation}).

%-----------------
\begin{figure}[!t]
\begin{center}
{\epsfig{figure=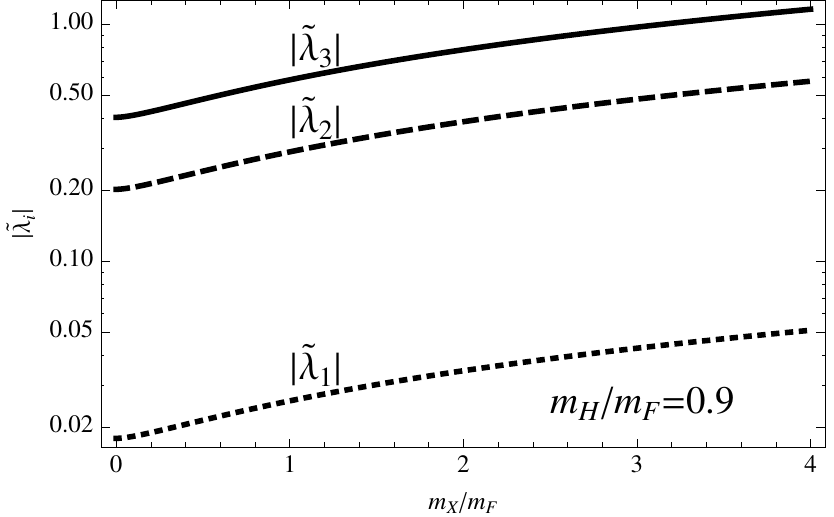,width=0.5\textwidth}}
\caption{The Yukawa couplings, $\widetilde \lambda_i$, depending on $m_X/m_F$. $m_H/m_F$ and $A/m_F$ are fixed at $0.9$ and $1$, respectively. $|\widetilde \kappa_2|=|\widetilde \lambda_2|$ and $|(V_{R})_{ij}|= |V_{ij}|$ are assumed. Those assumptions lead $|\widetilde \lambda_i|=|\widetilde \kappa_i|$. }
\label{fig;Yukawa-HF01}
\end{center}
\end{figure}
%------------------------------

Assuming $V_R \simeq V$ and using the relation of the diagonal elements in Eq. (\ref{eq;relation}), $y^d_i$ can be approximately estimated as 
\beq
\label{eq;diagonal}
|y^d_1|  = {\cal O} \left(\sqrt 2 m_d/v \right ),~| y^d_2| = {\cal O} \left(\sqrt 2 m_s/v \right ),~y^d_3  \simeq \frac{\sqrt 2}{v} \, m_b.
\eeq 
As discussed below, $\epsilon \widetilde \kappa_2 \widetilde \lambda_2$ becomes the same order as $\sqrt 2  m_s/v$, but
$|y^d_2|={\cal O}(\sqrt 2 m_s/v)$ is also required to be realistic. 

Similarly, the off-diagonal elements of Eq. (\ref{eq;relation}) lead the conditions for $\widetilde \kappa_i$ and $\widetilde \lambda_i$.
The assumption, $V_R \simeq V$, implies $\widetilde \kappa_i \widetilde \lambda_j \simeq \widetilde \kappa_j \widetilde \lambda_i$  and each size is estimated as
\beq
|\widetilde \kappa_1 \widetilde \lambda_2| =  {\cal O}(V_{us}) , ~ |\widetilde \kappa_1 \widetilde \lambda_3| = {\cal O}\left(|V_{ub}|\, m_b/m_s \right ), ~ |\widetilde \kappa_2 \widetilde \lambda_3| = {\cal O}\left(V_{cb}\, m_b/m_s \right ).
\eeq
Assuming $|\widetilde \kappa_i| \simeq |\widetilde \lambda_i|$, the alignment of $\widetilde \kappa_i $ and $\widetilde \lambda_i$ are expected to be
\beq
 (|\widetilde \kappa_1|, \,|\widetilde \kappa_2|, \, |\widetilde \kappa_3|)\simeq (|\widetilde \lambda_1|, \,|\widetilde \lambda_2|, \,|\widetilde \lambda_3|) \simeq  \left( \frac{|V_{ub}|}{|V_{cb}|}  , \, 1, \, \frac{|V_{ub}|}{|V_{us}|} \frac{|y^d_3|}{ |y^d_2|} \right ) \times |\widetilde \lambda_2|.
\eeq

In our numerical study discussed below, we approximately evaluate $\widetilde \kappa_i $ and $\widetilde \lambda_j$, assuming 
\beq
y^d_1 = \frac{m_d \sqrt 2  }{v} ,~y^d_3 = \frac{ m_b \sqrt 2}{v}.
\eeq
$y^d_2$ is fixed by the (2, 2) element of Eq. (\ref{eq;relation}).
The off-diagonal elements of Eq. (\ref{eq;relation}) lead 
$\widetilde \kappa_i \widetilde \lambda_j$: to a good approximation,
\begin{eqnarray}
\widetilde \kappa_1 \widetilde \lambda_2 &=& -\frac{y^d_2}{\epsilon} V_{us} , ~\widetilde \kappa_2 \widetilde \lambda_1= -\frac{y^d_2}{\epsilon} |(V_{R})_{12}| e^{-i (\delta_R)_{12}},  \nonumber \\
\widetilde \kappa_1 \widetilde \lambda_3 &=& -\frac{y^d_3}{\epsilon} V_{ub} , ~\widetilde \kappa_3 \widetilde \lambda_1= -\frac{y^d_3}{\epsilon} |(V_{R})_{13}| \, e^{-i (\delta_R)_{12}-i (\delta_R)_{23}-i \beta}, \nonumber \\
\widetilde \kappa_2 \widetilde \lambda_3 &=& -\frac{y^d_3}{\epsilon} V_{cb} , ~\widetilde \kappa_3 \widetilde \lambda_2= -\frac{y^d_3}{\epsilon} |(V_{R})_{23}| \, e^{-i (\delta_R)_{23}}.\label{eq;relation0}
\end{eqnarray}
$\beta$ is from the CKM matrix: $V_{ub}=|V_{ub}|e^{-i \beta}$. Note that there is a prediction for $V_R$ and $y^d_i$
according to the mass matrix in Eq. (\ref{eq;mass matrix}):
\beq
\label{eq;relation-1}
\frac{Y_{31}}{Y_{13}}=\frac{Y_{21}}{Y_{12}}\frac{Y_{32}}{Y_{23}}
\eeq 
where $Y_{ij}=\left ( V y^d V^\dagger_R \right )_{ij}$ is defined. 
$(V_{R})_{ij}$ need satisfy this condition and
$|(V_{R})_{ij}|=|V_{ij}|$ is assumed in our study. Note that in this parametrization $y^d_2$ is evaluated as
\beq
y^d_2 = \frac{ m_s \sqrt 2}{v}\frac{V_{ub}}{V_{ub}-V_{us}V_{cb}}.
\eeq

This estimation is a good approximation to realize the observed Yukawa couplings.
We expect that there are also some corrections from other extra particles, such as $\phi_{1,2}$ and $H^\prime_{1,2}$, so we allow 10 \% deviation in the down-type quark masses.
In our analysis, we use the input parameters for the quark masses \cite{PDG} and the CKM matrix \cite{CKMfitter}
derived from the values in Table \ref{table;input}.
When we compare our prediction with the realistic Yukawa couplings,
we evaluate the quark masses at $1$ TeV, using the SM RG running at the two-loop level \cite{RunDec,SMrun}.

Figure \ref{fig;Yukawa-HF01} shows the size of $|\widetilde \lambda_i|$ depending on $m_X/m_F$,
assuming that $|\widetilde \kappa_2|=|\widetilde \lambda_2|$ and $|(V_{R})_{ij}|= |V_{ij}|$.
The CP phases and $m_H/m_F$ are fixed at $(\delta_R)_{12}=(\delta_R)_{23}=0$ and $m_H/m_F=0.9$, respectively.
It is interesting that this hierarchical Yukawa couplings, $\widetilde \lambda_i$, 
have been proposed in Ref. \cite{Okawa}, motivated by DM physics and flavor physics.

In Sec. \ref{sec3}, we discuss flavor physics in this setup and 
specify the DM mass region consistent with the experimental results. We see that  $ \widetilde \kappa_{i} \widetilde \lambda_j$, which appear in Eq. (\ref{eq;relation}), directly relate to the flavor violating couplings which contribute to the $\Delta F=2$ processes.  Therefore, we can explicitly derive the lower bound on the flavor symmetry breaking scale.

%%%%%%%%Input parameters%%%%%%%%%%%%%%%%
\begin{table}
\begin{center}
  \begin{tabular}{|c|c||c|c|} \hline
 $m_d$(2 GeV) & 4.8$^{+0.5}_{-0.3}$ MeV \cite{PDG}  &  $\lambda$& 0.22509$^{+0.00029}_{-0.00028}$  \cite{CKMfitter}     \\ 
    $m_s$(2 GeV) & 95$\pm 5$ MeV \cite{PDG} & $A$& $0.8250^{+0.0071}_{-0.0111}$  \cite{CKMfitter}  \\ 
      $m_{b}(m_b)$&4.18$\pm 0.03$ GeV  \cite{PDG}    &  $\overline{\rho}$& 0.1598$^{+0.0076}_{-0.0072}$  \cite{CKMfitter} \\ 
  $\frac{2m_{s}}{(m_u + m_d)}$(2 GeV)& 27.5$\pm 1.0$ \cite{PDG}   &  $\overline{\eta}$&   0.3499$^{+0.0063}_{-0.0061}$ \cite{CKMfitter} \\ 
     $m_{c}(m_c)$&1.275$\pm 0.025$ GeV  \cite{PDG}  & $M_Z$ & 91.1876(21) GeV  \cite{PDG}  \\ 
    $m_t(m_t)$& 160$^{+5}_{-4}$ GeV  \cite{PDG}  &  $M_W$ & $80.385(15)$ GeV  \cite{PDG}   \\ 
 $\alpha$ & 1/137.036  \cite{PDG}  &  $G_F$  & 1.1663787(6)$\times 10^{-5}$ GeV$^{-2}$  \cite{PDG}       \\ 
 $\alpha_s(M_Z)$ & $0.1193(16)$ \cite{PDG}   &  &
 % $\sin^2 \theta_W$ & 0.23126(5)  \cite{PDG} 
   \\ \hline
  \end{tabular}
 \caption{The input parameters in our analysis. The CKM matrix, $V$, is written in terms of $\lambda$, $A$, $\overline{\rho}$ and $\overline{\eta}$ \cite{PDG}.}
  \label{table;input}
  \end{center}
\end{table}
%%%%%%%%%%%%%%%%%%%%%%%%%%%%%%%%

%%%%%%%%%%%%%%%%%%%%%%%%%%%%
\section{Flavor physics}
\label{sec3}
In this section, we discuss flavor physics in our scenario.
As well known, the $\Delta F=2$ processes, such as $K_0$-$\overline{K_0}$,
are the most sensitive to the new physics contributions.
In addition, the CP-violation and the rare meson decay possibly constrain our model.
First, we study the $\Delta F=2$ processes and discuss 
the deviations from the SM predictions, based on the result in Sec. \ref{sec2}.
Then, we study the other observables: e.g., $B \to X_s \, \gamma$ and
the neutron EDM. In particular, we see that
our model is strongly constrained by the CP-violation 
of the $K_0$-$\overline{K_0}$ mixing and the EDM.

 %%%%%%%%%%%%%%%%%%%%%%%%%%%%
\subsection{$\Delta F=2$ processes}
In our scenario, the CKM matrix is radiatively generated by $F$, $X$ and $H_D$ in the hidden sector.
In addition, the extra fields induce the operators relevant to the $\Delta F=2$ processes:
\begin{eqnarray}
{\cal H}^{\Delta F=2}_{eff}&=&( C_1)_{ij} (\overline{d^i_L} \gamma^\mu d^j_L)  (\overline{d^i_L} \gamma_\mu d^j_L) + (\widetilde C_1)_{ij} (\overline{d^i_R} \gamma^\mu d^j_R)  (\overline{d^i_R} \gamma_\mu d^j_R)   \nonumber \\
&&+( C_{4})_{ij} (\overline{d^i_L} d^j_R)  (\overline{d^i_R}  d^j_L) +h.c.. \label{eq;deltaF2}
\end{eqnarray}
The Wilson coefficients at the one-loop level are given by,
\begin{eqnarray}
(C_1)_{ij}&=& \frac{( \widetilde \kappa_i \widetilde \kappa^*_j    )^2}{64 \pi^2 m^{ 2}_{F}}   {\cal B}_1(m^2_H/m^2_F) , \\
(\widetilde C_1)_{ij}&=& \frac{( \widetilde \lambda^*_i \widetilde \lambda_{ j}     )^2}{64 \pi^2 m^{ 2}_{F}}  {\cal B}_1(m^2_X/m^2_F), \\
(C_{4})_{ij}&=&-\frac{ \widetilde \kappa_i  \widetilde \lambda_j  \widetilde \kappa^*_j  \widetilde \lambda^*_i }{16 \pi^2 m^2_F} \, {\cal B} (m^2_H/m^2_F,m^2_X/m^2_F),
\end{eqnarray}
where ${\cal B}_1 (x,y)$ and ${\cal B} (x,y)$ are defined as
\begin{eqnarray}
{\cal B}_1(x)&=&\frac{1}{(1-x)^2}
\left ( \frac{1+x}{2} + \frac{x}{1-x} \ln x  \right ),  \\
{\cal B} (x,y)&=& \frac{-x (y-1)^2 \ln x+ y (x-1)^2 \ln y-(x-1)(y-1)(x-y)}{(x-1)^2(y-1)^2(x-y)}.
\end{eqnarray}
As shown in Fig. \ref{fig;Yukawa-HF01}, the Yukawa couplings, $\widetilde \kappa_i$ and $\widetilde \lambda_i$,
are sizable in our models, so that the constraints from the $K_0$-$\overline{K_0}$ mixing 
should be taken into account, because it is the most sensitive to new physics among the observables of the $\Delta F=2$ processes. 

%%%%%%%%Input parameters%%%%%%%%%%%%%%%%
\begin{table}
\begin{center}
  \begin{tabular}{|c|c||c|c|} \hline
    $m_K$ & 497.611(13) MeV  \cite{PDG}  & $m_{B_s}$ & 5.3663(6) GeV \cite{PDG}  \\ 
     $F_K$ & 155.8(17) MeV \cite{Lattice1} & $m_{B}$ & 5.2795(3) GeV \cite{PDG} \\ 
     $\Hat B_K$ & 0.7625(97) \cite{Lattice2}  &  $f_{B_s} \sqrt{\Hat B_{B_s}}$ & 270(16) MeV \cite{Lattice2} \\ 
  $(\Delta M_K)_{\rm exp}$  & 3.484(6)$\times 10^{-12}$ MeV  \cite{PDG} & $f_{B} \sqrt{\Hat B_{B}}$ &219(14) MeV \cite{Lattice2}   \\ 
   $|\epsilon_K|_{\rm exp}$  & $2.228(11)\times 10^{-3}$  \cite{PDG}  &  $\Hat B_{B_s}$ & 1.32(6) \cite{Lattice2} \\ 
 $\eta_1$ & 1.87(76) \cite{Brod:2011ty}     & $\Hat B_{B}$ & 1.26(9)  \cite{Lattice2}  \\ 
    $\eta_2$& 0.5765(65) \cite{Buras:1990fn}      & $\eta_B$ & 0.55 \cite{Buras:1990fn}  \\ 
   $\eta_3$ & 0.496(47) \cite{Brod:2010mj}   & $\eta_Y$ &1.012 \cite{Buchalla:1998ba} \\  \hline
  \end{tabular}
 \caption{The input parameters relevant to the $\Delta F=2$ processes.}
  \label{table;input2}
  \end{center}
\end{table}
%   $|\epsilon_K|$ & $(2.228(11)) \times 10^{-3}$  \cite{PDG}
%%%%%%%%%%%%%%%%%%%%%%%%%%%%%%%%

In the $K$ system, we concentrate on $\epsilon_K$ and $\Delta M_K$.
They are approximately evaluated as
\beq
\epsilon_K= \frac{\kappa_\epsilon e^{i \varphi_\epsilon} }{\sqrt{2} (\Delta M_K)_{\rm exp}} \, Im(M^K_{12}), ~ \Delta M_K =2  Re(M^K_{12}),
\eeq
where $\kappa_\epsilon$ and $\varphi_\epsilon$ are $\kappa_\epsilon=0.94 \pm 0.02$ and $\varphi_\epsilon=0.2417 \times \pi$. $(\Delta M_K)_{\rm exp}$ is the experimental value and $M^K_{12}$ includes both the SM contribution and our prediction:
\begin{eqnarray}
M^{K \, *}_{12}&=&\left ( M^K_{12} \right )^*_{\rm SM}+ \left \{ ( C_1)_{sd} + ( \widetilde{ C}_1)_{sd} \right \} \times \frac{1}{3} m_K F^2_K \Hat B_K + ( C_4)_{sd}  \times \frac{1}{4}  \left ( \frac{m_K}{m_s + m_d} \right )^2 m_K F^2_K B_4. \nonumber \\
&&
\end{eqnarray}
The first term is the SM prediction described by 
$(M^K_{12})_{\rm SM}$,
\beq
(M^K_{12})_{\rm SM}^*= \frac{G^2_F}{12 \pi^2} F^2_K \Hat{B}_K m_K M^2_W \left \{ V^2_c \eta_1S_0( x_c) +  V^2_t \eta_2 S_0(x_t) + 2  V_c V_t \eta_3 S(x_c, x_t)  \right \},
\eeq
where $x_i$ and $V_i$ denote $(m^u_i)^2/M^2_W$ and $V^*_{is} V_{id}$, respectively. $\eta_{1,2,3}$ correspond to the NLO and NNLO QCD corrections. The values of our input parameters are summarized in Table \ref{table;input2}.
In our numerical analysis of the predictions, we use the central values.
Note that the the central values of the input parameters for the CKM matrix give $|V_{cb}|=41.80 \times 10^{-3}$
and $|V_{ub}|=3.71 \times 10^{-3}$. For these matrix elements,
it is known that there are discrepancies between the values derived from the exclusive and inclusive $B$ decay.
Each value is close to $|V_{cb}|$ ($|V_{ub}|$) of the inclusive (exclusive) decay, respectively. 
$B_K$ and $B_4$ are the bag parameters that are derived from the lattice calculation \cite{Lubicz:2008am}.
Of interest is to note that the $( C_4)_{sd} $ contribution, which directly relates to the Yukawa couplings
of the down-type quarks in Eq. (\ref{eq;mass matrix}), dominates over the other contributions.
To evaluate the Wilson coefficient, $( C_4)_{ij}$, at ${\cal O}(1)$ GeV,
we include the renormalization group (RG) correction at the one-loop level.

%-----------------
\begin{figure}[!t]
\begin{center}
{\epsfig{figure=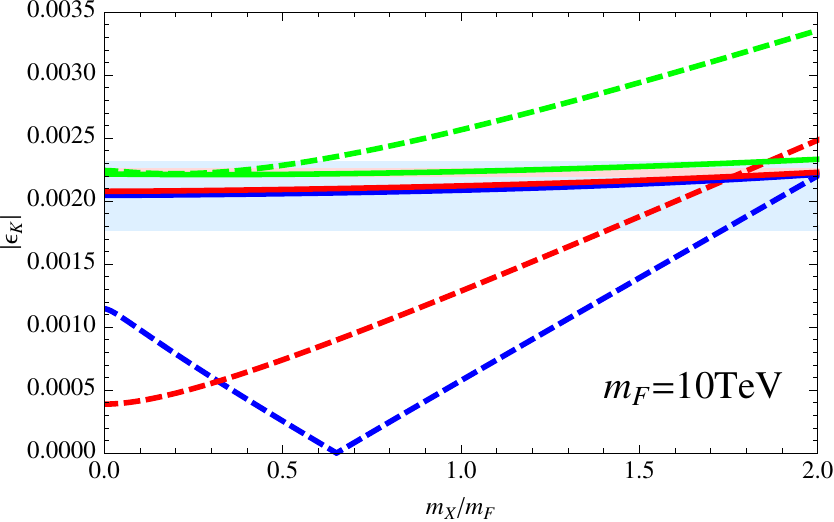,width=0.5\textwidth}}{\epsfig{figure=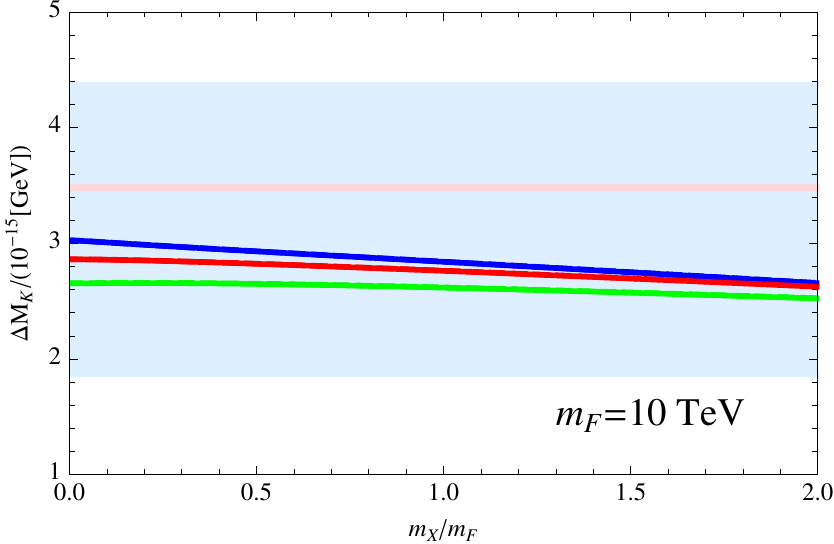,width=0.5\textwidth}}
\caption{ $m_X/m_F$ vs. $|\epsilon_K|$ (left panel) and $\Delta M_K$ (right panel). $m_F$ is fixed at $m_F=10$ TeV. Each line corresponds to $m_H/m_F=0.1$ (blue), $0.9$ (red), and $2$ (green).
The phases of $(V_R)_{ij}$ are fixed at $(\delta_R)_{23}=0$, $(\delta_R)_{12}=0$ (thick line) and $0.1$ (dashed line).
The light blue band is the SM prediction with 1$\sigma$ errors of $\eta_{1,2,3}.$
The pink band corresponds to $|\epsilon_K|_{\rm exp}$. }
\label{fig;mF10TeV-K}
\end{center}
\end{figure}
%------------------------------

Setting $m_F=10$ TeV and $m_H/m_F=0.1$ (blue), $0.9$ (red), $2$ (green), 
we draw our predictions of $|\epsilon_K|$ (left panel) and $\Delta M_K$ (right panel)  in Fig. \ref{fig;mF10TeV-K}.
The new phases in $(V_R)_{ij}$ are fixed at $(\delta_R)_{23}=0$, $(\delta_R)_{12}=0$ (thick line) and $0.1$ (dashed line), respectively. 
Using the central values in Table \ref{table;input2}, $|\epsilon_K|$ of the SM prediction is estimated as $|\epsilon_K|_{\rm SM}=2.04 \times10^{-3}$, that is slightly smaller than the experimental result: $|\epsilon_K|_{{\rm exp}}=2.228(11) \times 10^{-3}$  \cite{PDG}.
The SM prediction, however, suffers from the large uncertainty, so that it may be difficult to draw the
explicit exclusion limit. As we see Table \ref{table;input2},
the contributions involving charm quark has large errors, so that more than 10 \% ambiguity still exists even in $| \epsilon_K|$ of the SM prediction. 
The light blue bands on both panels are the SM predictions with 1$\sigma$ errors of $\eta_{1,2,3}.$
The pink bands correspond to $|\epsilon_K|_{\rm exp}$ and $|\Delta M_K|_{\rm exp}$, respectively.

If we require the deviation of $|\epsilon_K|$ from the SM prediction to be within the error, 
there is an allowed parameter region in the $m_F=10$ TeV case. 
The CP phase, $(\delta_R)_{12}$, is relevant to $|\epsilon_K|$, so that vanishing $(\delta_R)_{12}$ can evade the strong bound from the observable. 
As shown in Fig. \ref{fig;mF10TeV-K}, $|(\delta_R)_{12}|$ should be smaller
than $0.1$, unless $m_H$ is larger than $m_F$. 

\begin{figure}[!t]
\begin{center}
{\epsfig{figure=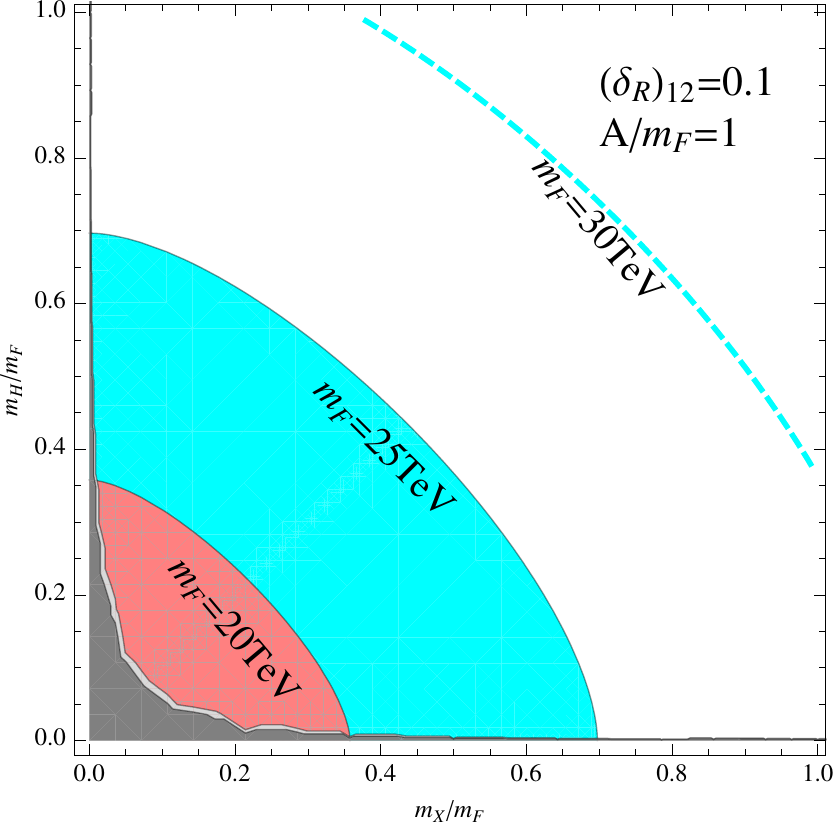,width=0.5\textwidth}}
\caption{The region that $|\epsilon_K|$ is within $1.79 \times 10^{-3}\leq |\epsilon_K| \leq 2.30 \times 10^{-3}$ in the cases with $m_F=20$ TeV (pink) and $25$ TeV (cyan). $(\delta_R)_{12}$, $(\delta_R)_{23}$ and $A/m_F$ are fixed at $((\delta_R)_{12},\, (\delta_R)_{23},\,A/m_F) =(0.1,\, 0,\,1)$. The dashed cyan line corresponds to the exclusion limit of
 $m_F=30$ TeV case.
The (light) gray region is excluded by the vacuum stability when $m_F$=20 TeV (25 TeV) and $A/m_F=1$ are satisfied.  }
\label{fig;contour}
\end{center}
\end{figure}

In Fig. \ref{fig;contour}, we draw the region that the deviation of $|\epsilon_K|$ from the SM prediction
is within the error of the SM prediction: $1.79 \times 10^{-3}\leq |\epsilon_K| \leq 2.30 \times 10^{-3}$. 
$(\delta_R)_{12}$, $(\delta_R)_{23}$ and $A/m_F$ are fixed at $((\delta_R)_{12},\, (\delta_R)_{23},\,A/m_F) =(0.1,\, 0,\,1)$. In the pink (cyan) region, the deviation is within the error in the case with $m_F=20$ TeV (25 TeV).
The exclusion limit reaches the dashed cyan line, when $m_F = 30$ TeV.
The (light) gray region in Fig. \ref{fig;contour} is excluded by the vacuum stability in Eq. (\ref{vacuum}) when $m_F=20$ TeV (25 TeV) is satisfied. 

$H_D$ or $X$ is a DM candidate, so that 
this figure shows the DM mass region, depending on $m_F$. For instance, both of $H_D$ 
and $X$ are lighter than $F$, when $m_F$ is below 25 TeV. On the other hand, 
either $H_D$ or $X$ could be heavier than $F$, if $m_F$ is 30 TeV or heavier.
Note that the constraint from $|\epsilon_K|$ is drastically relaxed, when $m_F$ is larger than 100 TeV.

%-----------------
%------------------------------

\begin{figure}[!t]
\begin{center}
{\epsfig{figure=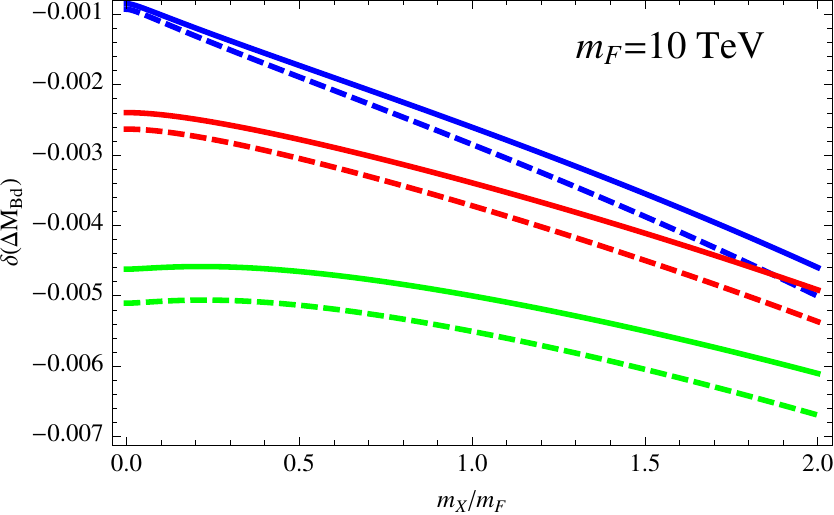,width=0.5\textwidth}}{\epsfig{figure=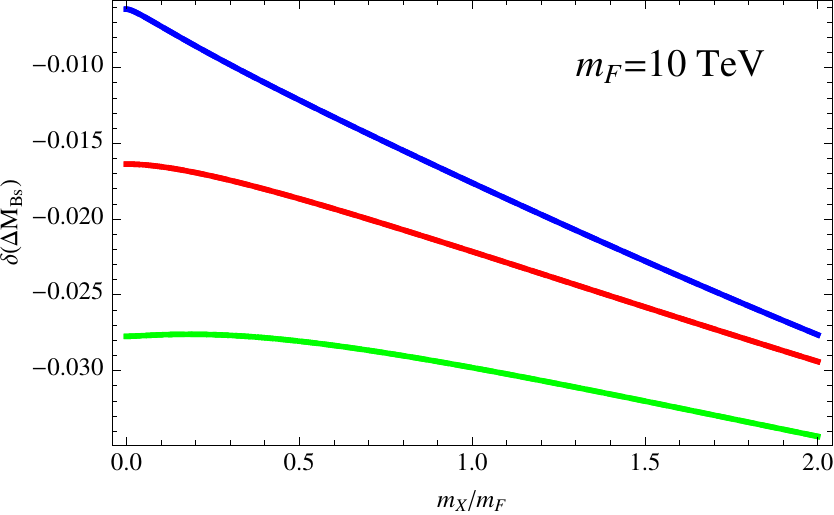,width=0.5\textwidth}} 
\caption{The deviations of $\Delta M_{B_d}$ and $\Delta M_{B_s}$ with $m_F=10$ TeV. Each line corresponds to $m_H/m_F=0.1$ (blue), $0.9$ (red), and $2$ (green).
The phases of $(V_R)_{ij}$ are fixed at $(\delta_R)_{23}=0$, $(\delta_R)_{12}=0$ (thick line) and $0.1$ (dashed line). }
\label{fig;mF10TeV-B}
\end{center}
\end{figure}

In the same manner, we can evaluate the $B_d$-$\overline{B_d}$ and $B_s$-$\overline{B_s}$ mixing.
The mass differences of the $B$ mesons in our model are given by
\beq
\Delta M_{B_q}= 2 \left |  M^{B_q}_{12} \right |^2 =2 \left | (M^{B_q}_{12})_{\rm SM} +\Delta M^{B_q}_{12} \right |^2 ~(q=d, \, s).
\eeq
$\Delta M^{B_q}_{12}$ describes the contributions of $(C_1)_{bq}$, $(\widetilde C_1)_{bq}$ and $(C_4)_{bq}$ in Eq. (\ref{eq;deltaF2}). 
Figure \ref{fig;mF10TeV-B} shows the deviations of $\Delta M_{B_d}$ (left panel) and $\Delta M_{B_s}$ (right panel),
when $m_F$ is fixed at 10 TeV. The parameter choice is the same as in Fig. \ref{fig;mF10TeV-K}.
$\delta ( \Delta M_{B_q}) = \Delta M_{B_q}/ ( \Delta M_{B_q})_{\rm SM} -1$ are defined in Fig. \ref{fig;mF10TeV-B}.

Using the central values in Table \ref{table;input2}, $\Delta M_{B_d}$ and $\Delta M_{B_s}$ of the SM predictions are estimated as $\Delta M_{B_d}=0.517$ [ps$^{-1}$] and $\Delta M_{B_s}=18.358$  [ps$^{-1}$]. The experimental results are $\Delta M_{B_d}=0.554^{+0.035}_{-0.028}$ [ps$^{-1}$] and $\Delta M_{B_s}=16.89^{+0.47}_{-0.35}$  [ps$^{-1}$]   \cite{CKMfitter}, respectively.
In this sense, $\delta ( \Delta M_{B_d} )$ should be positive and $\delta ( \Delta M_{B_d} )$ should be negative,
although the errors of $f_{B_q} \sqrt{\Hat B_{B_q}}$ in Table \ref{table;input2} cause about 10 \% uncertainties for the SM predictions.

We see that the deviations are less than 1 \% in $\Delta M_{B_q}$. There is a small dependence of 
$(\delta_R)_{12}$ in $\Delta M_{B_d}$, but the predicted deviation is not so large as far as $m_F$ is larger than
10 TeV.
In $\Delta M_{B_s}$, the deviation is relatively large, compared to $\delta (\Delta M_{B_d})$.
This is because $|\widetilde \lambda_{2}|$ is about 10 times larger
than $|\widetilde \lambda_{1}|$. Depending on the scalar masses, $|\delta (\Delta M_{B_s})|$ could reach
${\cal O}(0.01)$.

%-----------------
%------------------------------

\subsection{$B \to X_s \gamma$ and EDM}
We have seen that the strongest constraint on our model is from $|\epsilon_K|$.
In addition, we can find other flavor violating and CP-violating processes relevant to our scenario.
For instance, it is known that the rare $B$ meson decay, $B \to X_s \gamma$,
strongly constrains new physics contribution. In our scenario,
the one-loop diagram involving the mediators contributes to
the process.
The effective operators are given by
\beq
{\cal H}^{b \to s \gamma}_{eff}=-g_{{\rm SM}} \left \{ C_7 (\overline{s_L} \sigma^{\mu \nu} b_R) F_{\mu \nu}+C^\prime_7 (\overline{s_R} \sigma^{\mu \nu} b_L) F_{\mu \nu} \right \},
\eeq  
where $g_{{\rm SM}}$ is defined as
\beq
g_{{\rm SM}} =\frac{4 G_F}{\sqrt 2} (V^*_{ts}V_{tb}) \times \frac{e \, m_b}{ 16 \pi^2}.
\eeq
In our model, $C_7$ and $C^\prime_7$ are predicted as
\begin{eqnarray}
C_7&=& \frac{g^\prime}{ g_{{\rm SM}} }  \frac{ \widetilde \kappa_2 \widetilde \lambda_3 }{96 \pi^2 m^3_F}  A \frac{v}{\sqrt 2} \, {\cal B} (m^2_H/m^2_F,m^2_X/m^2_F), \\
C^\prime_7&=& \frac{g^\prime}{ g_{{\rm SM}}}  \frac{ \widetilde \kappa^*_3 \widetilde \lambda^*_2 }{96 \pi^2 m^3_F} A^* \frac{v}{\sqrt 2} \, {\cal B} (m^2_X/m^2_F,m^2_H/m^2_F),
\end{eqnarray}
where $g^{\prime}$ is the gauge coupling of U(1)$_Y$.
$|\epsilon_K|$ requires at least ${\cal O}(10)$-TeV colored particle,
so that $C_7$ and $C^\prime_7$ are suppressed by $v/m^2_F$.
Fixing $m_F=10$ TeV, we estimate those coefficients as
$C_7 \simeq {\cal O}(10^{-3})$. Such a small parameter predicts at most
a few \% deviation of Br($B \to X_s \gamma$), so that 
we conclude that the branching ratio including the new physics contribution is consistent with the combined experimental result: Br($B \to X_s \gamma$)=$(3.43 \pm 0.22)\times 10^{-4}$~\cite{Amhis:2014hma}.

The penguin diagrams also arise and contribute to $\Delta F=1$ processes.
In our model, those contributions are, however, suppressed by $(A v/m^2_X)^2$ 
or $(A v/m^2_H)^2$, that correspond to the mixing between $X$ and $H_D$.
Then, we can not expect large deviations in $\Delta F=1$ processes
through the penguin diagrams.

%-----------------
\begin{figure}[!t]
\begin{center}
{\epsfig{figure=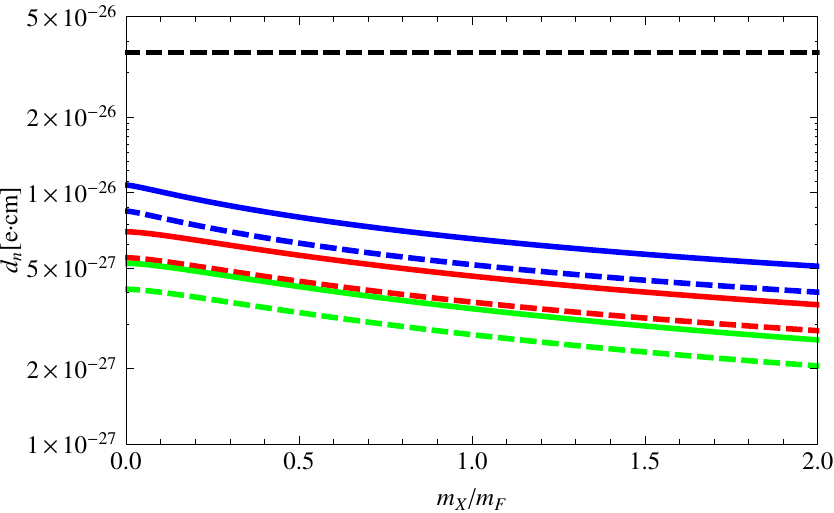,width=0.6\textwidth}}
\caption{The neutron EDM with $m_F=10$ TeV. Each line corresponds to $m_H/m_F=0.1$ (blue), $0.9$ (red), $2$ (green).
The phases of $(V_R)_{ij}$ are fixed at $(\delta_R)_{23}=0$, $(\delta_R)_{12}=0$ (thick line) and $0.1$ (dashed line). The dashed black line corresponds to the current exclusion line: $|d_n| <3.6 \times 10^{-26}$ [$e \, cm$] \cite{EDM}.}
\label{fig;mF10TeV-EDM}
\end{center}
\end{figure}
%------------------------------

Finally, we discuss electric dipole moments (EDMs) in our model. $\widetilde \kappa_i$ and $\widetilde \lambda_i$, in general, have non-vanishing imaginary parts, because of the CP-violating phases of the CKM matrix and $(V_R)_{ij}$.
The important point is that the relation in Eq. (\ref{eq;relation}) limits the phase, as shown in Eq. (\ref{eq;relation0}).
Then, we find that there is no parameter choice such that the CP-violating phases contributing to $\epsilon_K$ and EDMs cancel out at the same time.
We plot our prediction of the neutron EDM, $d_n$, that is constrained as $|d_n| < 3.6 \times 10^{-26}$ [$e \, cm$] \cite{EDM}. 
A lot of efforts have been done to improve the theoretical prediction \cite{Hisano:2012sc,Fuyuto:2012yf,Jung:2013hka}.
We adopt the value in Ref. \cite{Jung:2013hka} and draw our prediction in Fig. \ref{fig;mF10TeV-EDM}.
$m_F$ is fixed at $m_F=10$ TeV. Each line corresponds to $m_H/m_F=0.1$ (blue), $0.9$ (red), and $2$ (green).
The phases of $(V_R)_{ij}$ are fixed at $(\delta_R)_{23}=0$, $(\delta_R)_{12}=0$ (thick line) and $0.1$ (dashed line), respectively. The dashed black line corresponds to the current exclusion limit: $|d_n| < 3.6 \times 10^{-26}$ [$e \, cm$] \cite{EDM}. Our predictions are below the current experimental bound.
Note that our prediction for $|d_n|$ becomes smaller when $(\delta_R)_{12}$ is about $0.45$; on the other hand,  
$|\epsilon_{K}|$ becomes larger than in the $(\delta_R)_{12} \simeq 0$ case.

The measurement of the permanent EDM of neutral $^{199}H_g$ atom is developed and the current upper bound reaches 
$|d_{H_g}| < 7.4 \times 10^{-30}$ [$e \, cm$] \cite{permanentEDM}.
This measurement, however, suffers from the large uncertainty of the theoretical prediction \cite{Jung:2013hka}, so that
it is still difficult to compare our prediction with the experimental bound.
If we use the central values introduced in Ref. \cite{permanentEDM}, our prediction estimated as ${\cal O}(10^{-29})$ [$e \, cm$] when $m_F=10$ TeV, so that our model could be tested if the theoretical error is shrunken.

\section{Dark matter physics}
\label{sec4}
We have obtained the mass spectrum of the extra particles and the couplings between
the extra particles and quarks, according to the realistic quark mass matrix and the flavor physics.
Finally, we discuss DM physics in this section.

The DM candidate in our model is either $X$ or the neutral component of $H_D$,\footnote{We do not consider the case that both $X$ and the neutral component of $H_D$ are stable.} and both have couplings with the SM Higgs in the scalar potential as,
\beq
\label{scalar-int}
\lambda_X |X|^2 |H|^2 + \lambda_3 |H_D|^2 |H|^2+\lambda_4 |H^\dagger_DH|^2 ,
\eeq
 in addition to the trilinear coupling, $A X H^\dagger_D H$, and the Yukawa couplings with down-type quarks.
This type of DM has been studied recently \cite{Okawa}.
The authors of Ref. \cite{Okawa} concentrate on the relatively light $F$ case,
and do the integrated research of the LHC physics, the flavor physics, and the DM physics.
In our scenario, $F$ should be at least ${\cal O}(10)$ TeV, to avoid too large deviations
in the $K_0$-$\overline{K_0}$ mixing and the EDM, so that our parameters are
out of the region analyzed in Ref. \cite{Okawa}.

When $m_F$ is ${\cal O}(10)$ TeV and $A/m_F$ is set to unit, 
the condition for the vacuum stability in Eq. (\ref{vacuum}) leads the DM mass region as
\beq
\label{vacuum2}
\frac{m_X m_{H}}{m^2_F} \gtrsim {\cal O} (10^{-2}).
\eeq
If we assume that there is no large hierarchy between $m_X$ and $m_{H}$,
this inequality means $m_X$ and $m_{H}$ should be not less than ${\cal O}(1)$ TeV.

The DM is thermally produced by the interactions with the SM Higgs in Eq. (\ref{scalar-int}) and  
the Yukawa interactions with the down-type quarks and $F$.
As shown in Fig. \ref{fig;Yukawa-HF01}, the alignment of the Yukawa couplings is
hierarchical, so that the annihilation of the DM to the bottom quarks is relatively larger
in the $t$-channel $F$ exchanging processes. The heavy $F$ mass of ${\cal O}(10)$-TeV, however, suppresses the 
annihilation, so that the main annihilation process is given by the
interaction with the SM Higgs in Eq. (\ref{scalar-int}).
In order to achieve the observed relic abundance of DM, DM mass should be less than ${\cal O}(10)$ TeV,
to respect the perturbativity of $\lambda_X$ \cite{Higgsportal}.\footnote{Note that we can also derive the upper bound of DM mass from the unitarity of the annihilation cross section; that is, the upper bound is  ${\cal O}(100)$ TeV \cite{Griest:1989wd}.}
Assuming that $X$ is DM, $m_X$ should be in the range,
\beq
\label{eq;mxrange}
{\cal O}(1) \, {\rm TeV}  \lesssim m_X \lesssim  10 \, {\rm TeV}.
\eeq
The upper bound comes from $\lambda_X \leq \sqrt{4 \pi}$ and the lower bound corresponds to Eq. (\ref{vacuum2}).
In this region, we can estimate the cross section for the direct detection.
The dominant process is the SM Higgs exchanging, and the cross section is
almost fixed once $\lambda_X$ is given by the thermal relic density.
The prediction of the spin-independent (SI) cross section with the nucleon at the direct detection experiments is
\beq
\label{eq;SI}
\sigma_{\rm SI} \simeq1.7 \times 10^{-9} \, [{\rm pb}].
\eeq
This prediction slightly depends on the DM mass, but the change is 
only a few percent as far as $m_X$ is within the mass region in Eq. (\ref{eq;mxrange}).
Note that there is a small correction from the $F$ exchanging in the SI cross section, Eq.(\ref{eq;SI}), 
but we find that it is not more than 10\% of the Higgs exchanging contribution when $m_F$ is ${\cal O}(10)$ TeV. 
The current upper bound is given by the LUX and the PandaX-II experiments: $\sigma_{\rm SI} \lesssim {\cal O}( 10^{-8}) \, [{\rm pb}]$ \cite{LUX2015,LUX2016,Panda}.
 In the future, the XENON1T experiment could reach ${\cal O}(10^{-9})$ pb \cite{XENON1T}, so that
 our scenario is expected to be probed by the direct detection of DM.

In the case that the neutral component of $H_D$ is DM, the DM physics is more complicated
because the DM interacts with the SM particles via the $Z$ and $W$ gauge boson exchanging.
There are CP-even and CP-odd neutral scalars, and a charged scalar in $H_D$. 
The $\lambda_4$ term in the scalar potential generates the mass difference between the charged and neutral scalars, 
while this term does not split two neutral scalars. 
However, if the CP-even and the CP-odd scalars of $H_D$ are degenerate, the $Z$-boson exchanging contribution dominates
the cross section with the nuclei and the predicted $\sigma_{\rm SI}$ is excluded by the current experimental bound \cite{Higgsportal}. 
Therefore, the mass difference of two neutral scalars needs to be generated in this case. 
The mass splitting of the neutral scalars appears only when $\lambda_5 (H^\dagger_D H)^2$ is allowed in the potential. 
If the dark symmetry is (global) $U(1)$, the $\lambda_5$ term is forbidden, so that we conclude the dark symmetry should be a discrete $Z_2$ symmetry when we discuss the case that the neutral component of $H_D$ is DM.

In the annihilation processes, $H_D$ can annihilate to $Z$ and $W$ gauge bosons
as well as the SM fermions.
When the DM mass region of $H_D$ is not less than ${\cal O}(1)$ TeV, 
the main annihilation processes are the annihilations to the weak gauge bosons.
This kind of DM scenario has been studied well based on the recent experimental results \cite{IDM0,IDM0-1,IDM0-2,IDM0-3,IDM}.
In this scenario, the mass differences of the scalars depend on the couplings, $\lambda_4$ and $\lambda_5$.
We find that less than ${\cal O}(10)$ GeV mass differences among the scalars of $H_D$ can achieve the correct
relic density in the DM mass region given by Eq. (\ref{eq;mxrange}) \cite{IDM}. 
The cross section for the spin-independent direct detection is
estimated as Eq. (\ref{eq;SI}), so the $H_D$ DM could also be tested in the XENON1T experiment.

\section{Summary and Discussion}
\label{sec5}
%-----------------
\begin{figure}[!t]
\begin{center}
{\epsfig{figure=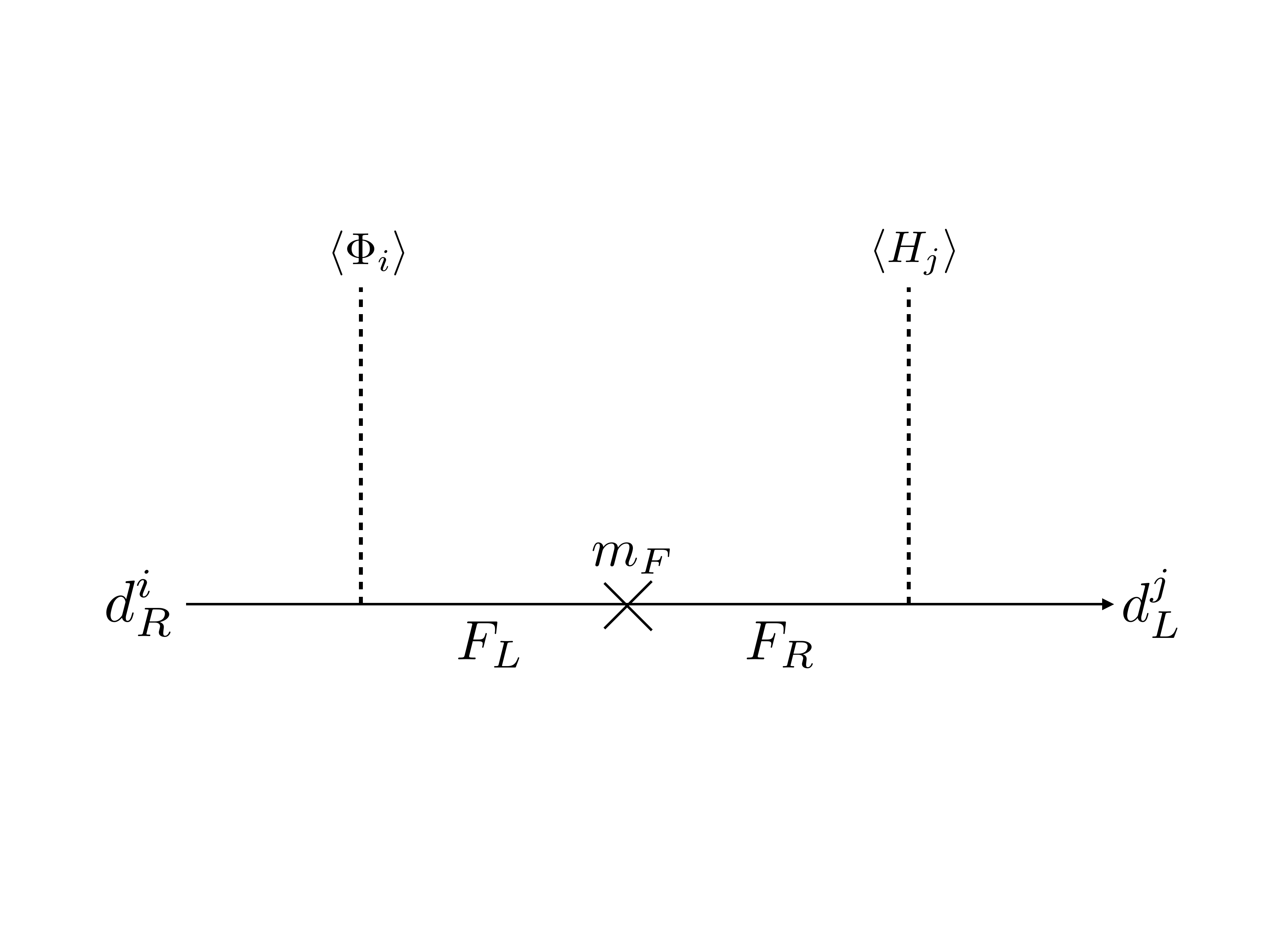,width=0.5\textwidth}}
\caption{Rough sketch of another idea to generate the flavor violation. }
\label{fig;idea2}
\end{center}
\end{figure}
%------------------------------
The flavor structure in our nature is one of mysteries, that may be revealed by the Beyond Standard Model. 
We do not know why the fermion masses are so hierarchical and the quark mixing is very small.
In the SM model, the flavor violating processes are described by the CKM matrix in the $W$ boson interaction,
and this description is consistent with the experimental results. 
The CKM matrix is very close to an identity matrix, but has small off-diagonal elements.
This corresponds to the different mass bases of up-type quarks and of down-type quarks,
so that this fact may imply the existence of new particles that interact with either up-type quarks
or down-type quarks.

In this paper, we propose the possibility that the flavor symmetry breaks down in the hidden sector existing around ${\cal O}(1)$ TeV-${\cal O}(10)$ TeV, 
and some extra particles mediate the flavor violating effect to the SM quark sector. Among the mediators, we can find DM candidates as well. The CKM mixing is radiatively generated, so that the CKM matrix directly relates to the structure of the mediation sector: the couplings with quarks and the masses of the mediators. We simply assume that the DM contribution to the CKM matrix is dominant, because DM is expected to be the lightest
particle among the mediators. Then, we derive the connection among the CKM matrix, the flavor physics and
the DM physics. Interestingly, the constraints from the vacuum stability, the flavor physics and the DM relic density require that the DM mass is between ${\cal O}(1)$ TeV and ${\cal O}(10)$ TeV, and the spin-independent DM scattering cross section is close to the expected region of the XENON1T experiment. We find the significant deviations in the flavor physics, so that we can test our scenario in the future experiments of the flavor physics as well.

Our main motivations are to find the connection between DM and the quark mixing in the SM
and to look for universal predictions of this kind of model that the CKM matrix is originated from the radiative corrections involving DM. Therefore, we have not constructed any explicit model for the hidden sector in this work, and
concentrate on the light-scalar contributions in the simple assumption. 
Our results could be applied to many concrete models that radiatively induce the quark mixing
and realize a DM candidate. The model-dependent analysis is, however, important
to understand how large the parameter region covered by our study is. 
For instance, the flavor physics has been studied in the minimal supersymmetric model,
where the quark mass matrix is radiatively induced \cite{Crivellin:2011sj}.
In the model, the one-loop diagrams involving the superpartners of gluon and quarks lead the
quark mixing and the mass hierarchy. Compared with our setup in Eq. (\ref{eq;relation}),
the bare coupling, $y^i_d$, has only one non-vanishing element and flavor U(2) symmetry is assigned in Ref. \cite{Crivellin:2011sj}.
In this case, the bound from the $K_0-\overline{K_0}$ mixing could be relaxed, 
if we assume that approximately only one U(2) breaking term, namely A-term, defines the mass eigenstates of the strange and down quarks. Then, one does not need large radiative contribution to realize the Cabibbo angle.
In our setup, on the other hand, all mass eigenstates are given by the linear combinations of the bare couplings and the radiative corrections and the Cabibbo angle is generated by the radiative correction.
Therefore, the bounds from the $K_0-\overline{K_0}$ mixing and the EDM, that are only relevant to the radiative corrections, cannot be evaded.  \footnote{We can also find the strong bound from the $K_0-\overline{K_0}$ mixing in the supersymmetric model \cite{Altmannshofer:2014qha,Ferrandis:2004ri}.} In addition, we suggest that the constraint on the CP-phase is the most stringent when the mass matrix is approximately in the form of Eq. (\ref{eq;relation}).
The study of the model with different forms from Eq. (\ref{eq;relation}) will be given in future \cite{Okawa2}.

Let us also discuss the other scenarios, motivated by the origin of the CKM matrix.
We did not qualitatively take into account the physics of $\phi_{1,2}$, $H^\prime_{1,2}$ and the fields
to break the flavor symmetry. There is a possibility that
tree-level diagrams, as in Fig. \ref{fig;idea2}, realize the quark mixing.
In this case, some of the scalars such as $\phi_{1,2}$ develop the non-vanishing VEVs
and the tree-level diagram simply generates the realistic down-type Yukawa couplings.
This kind of model is much simpler and has been discussed, for instance,
in the framework of the grand unified theory \cite{Barbieri:1980vc,Barbieri:1980tz,Barbieri:1981yw,Barbieri:1981yy,Balakrishna:1988ks,Hisano:2015pma,Hisano:2016afc}. 
In particular, the authors of Refs. \cite{Hisano:2015pma,Hisano:2016afc} recently consider such a simple setup
for the realistic Yukawa couplings in the SO(10) grand unified theory, and 
study the FCNCs predicted by the fermion mass hierarchies and the quark mixing.
In this case, however, scalar DM candidates, $X$ and $H_D$, may decay to quarks because of the non-vanishing VEVs of $\phi_{1,2}$, so that we may have to introduce some additional particles to realize a DM candidate.
Besides, there are tree-level FCNCs suppressed by the masses of the extra colored particles. 
In this scenario, the predicted mass matrix of down-type quarks is in the same 
form introduced in Eq. (\ref{eq;mass matrix}), so that we could also apply our analysis to 
this new scenario. Including this type of diagram in Fig. \ref{fig;idea2}, we will summarize possible setups motivated by both DM and the origin of the CKM matrix, and then discuss the universal predictions, the differences and relevant physics of each model \cite{Okawa2}.

%---------------------------------------------------------------------------

%%%%%%%%%%%%%%%%%%%%%%%%%%%%%%%%%%%%
\section*{Acknowledgments}
%%%%%%%%%%%%%%%%%%%%%%%%%%%%%%%%%%%%
%---------------------------------------------------------------------------
We would like to thank Alejandro Ibarra for discussions and suggestions. 
S.O. is also grateful to the hospitality of Physik-Department T30d, Technische Universit\"at M\"unchen where the first stage of this work was done. 
The work of Y. O. is supported by Grant-in-Aid for Scientific research from the Ministry of Education,
Science, Sports, and Culture (MEXT), Japan, No. 17H05404. 
%---------------------------------------------------------------------------


\begin{thebibliography}{99}

\bibitem{DMexperiment2}
  G. Hinshaw {\it et al.} [WMAP Collaboration],
  %``Nine-Year Wilkinson Microwave Anisotropy Probe (WMAP) Observations: Cosmological Parameter Results,''
  Astrophys.\ J.\ Suppl. 208:19 (2013)
%  doi:10.1051/0004-6361/201525830
  [arXiv:1212.5226 [astro-ph.CO]].
  %%CITATION = doi:10.1051/0004-6361/201525830;%%
  %2419 citations counted in INSPIRE as of 16 Nov 2016


%\cite{Ade:2015xua}
\bibitem{DMexperiment}
  P.~A.~R.~Ade {\it et al.} [Planck Collaboration],
  %``Planck 2015 results. XIII. Cosmological parameters,''
  Astron.\ Astrophys.\  {\bf 594}, A13 (2016)
%  doi:10.1051/0004-6361/201525830
  [arXiv:1502.01589 [astro-ph.CO]].
  %%CITATION = doi:10.1051/0004-6361/201525830;%%
  %2419 citations counted in INSPIRE as of 16 Nov 2016



%%%%%%%%%%%%%%%%%%%%%%%%%%%%%%%%
%%%%%%%%%%%%%%%%%%%%%%%%%%%%%%%%
%%%%%%%%radiative mass  generation%%%%%%%%%%%%%%%%%%%%
%%%%%%%%%%%%%%%%%%%%%%%%%%%%%%%

%%%%%%%% GUT  %%%%%%%%%%%


%\cite{Barbieri:1980vc,Barbieri:1980tz,Barbieri:1981yw}
\bibitem{Barbieri:1980vc} 
  R.~Barbieri and D.~V.~Nanopoulos,
  %``An Exceptional Model for Grand Unification,''
  Phys.\ Lett.\  {\bf 91B}, 369 (1980).
 % doi:10.1016/0370-2693(80)90998-3
  %%CITATION = doi:10.1016/0370-2693(80)90998-3;%%
  %209 citations counted in INSPIRE as of 01 Mar 2017
%\cite{Barbieri:1980tz,Barbieri:1981yw}


\bibitem{Barbieri:1980tz} 
  R.~Barbieri and D.~V.~Nanopoulos,
  %``Hierarchical Fermion Masses From Grand Unification,''
  Phys.\ Lett.\  {\bf 95B}, 43 (1980).
%  doi:10.1016/0370-2693(80)90395-0
  %%CITATION = doi:10.1016/0370-2693(80)90395-0;%%
  %76 citations counted in INSPIRE as of 01 Mar 2017
  %\cite{Barbieri:1981yw}
\bibitem{Barbieri:1981yw} 
  R.~Barbieri, D.~V.~Nanopoulos and D.~Wyler,
  %``Hierarchical Fermion Masses in SU(5),''
  Phys.\ Lett.\  {\bf 103B}, 433 (1981).
%  doi:10.1016/0370-2693(81)90076-9
  %%CITATION = doi:10.1016/0370-2693(81)90076-9;%%
  %38 citations counted in INSPIRE as of 01 Mar 2017
%\cite{Barbieri:1981yy}



\bibitem{Barbieri:1981yy} 
  R.~Barbieri, D.~V.~Nanopoulos and A.~Masiero,
  %``Hierarchical Fermion Masses in E6,''
  Phys.\ Lett.\  {\bf 104B}, 194 (1981).
  %doi:10.1016/0370-2693(81)90589-X
  %%CITATION = doi:10.1016/0370-2693(81)90589-X;%%
  %73 citations counted in INSPIRE as of 01 Mar 2017



  
%%%%%%%%%%%LR model %%%%%%%%%%%



%\cite{Kramer:1981sq,Balakrishna:1988ks,Balakrishna:1987qd}
\bibitem{Kramer:1981sq} 
  G.~Kramer and I.~Montvay,
  %``Radiative Quark Mass Generation And A Fourth Quark Family,''
  Z.\ Phys.\ C {\bf 11}, 159 (1981).
 % doi:10.1007/BF01573999
  %%CITATION = doi:10.1007/BF01573999;%%
  %4 citations counted in INSPIRE as of 01 Mar 2017
\bibitem{Balakrishna:1987qd} 
  B.~S.~Balakrishna,
  %``Fermion Mass Hierarchy From Radiative Corrections,''
  Phys.\ Rev.\ Lett.\  {\bf 60}, 1602 (1988).
  %doi:10.1103/PhysRevLett.60.1602
  %%CITATION = doi:10.1103/PhysRevLett.60.1602;%%
  %95 citations counted in INSPIRE as of 17 Feb 2017,
  %\cite{Balakrishna:1988ks}
\bibitem{Balakrishna:1988ks} 
  B.~S.~Balakrishna, A.~L.~Kagan and R.~N.~Mohapatra,
  %``Quark Mixings and Mass Hierarchy From Radiative Corrections,''
  Phys.\ Lett.\ B {\bf 205}, 345 (1988).
  %doi:10.1016/0370-2693(88)91676-0
  %%CITATION = doi:10.1016/0370-2693(88)91676-0;%%
  %112 citations counted in INSPIRE as of 17 Feb 2017
  
  

  %%%%%%%%%%%%SUSY%%%%%%%%%%%%%%%%%%%
  
  
  %\cite{Lahanas:1982et,Masiero:1983ph,Barr:1984pk,Kagan:1989fp}
\bibitem{Lahanas:1982et} 
  A.~B.~Lahanas and D.~Wyler,
  %``Radiative Fermion Masses and Supersymmetry,''
  Phys.\ Lett.\  {\bf 122B}, 258 (1983).
%  doi:10.1016/0370-2693(83)90696-2
  %%CITATION = doi:10.1016/0370-2693(83)90696-2;%%
  %29 citations counted in INSPIRE as of 01 Mar 2017
  
  %\cite{Masiero:1983ph,Barr:1984pk,Kagan:1989fp}
\bibitem{Masiero:1983ph} 
  A.~Masiero, D.~V.~Nanopoulos and K.~Tamvakis,
  %``Radiative Fermion Masses in Supersymmetric Theories,''
  Phys.\ Lett.\  {\bf 126B}, 337 (1983).
  %doi:10.1016/0370-2693(83)90176-4
  %%CITATION = doi:10.1016/0370-2693(83)90176-4;%%
  %16 citations counted in INSPIRE as of 01 Mar 2017
  
  %\cite{Barr:1984pk,Kagan:1989fp}
\bibitem{Barr:1984pk} 
  S.~M.~Barr,
  %``Light Fermion Mass Hierarchies in Supersymmetric Models,''
  Phys.\ Rev.\ D {\bf 31}, 2979 (1985).
 % doi:10.1103/PhysRevD.31.2979
  %%CITATION = doi:10.1103/PhysRevD.31.2979;%%
  %12 citations counted in INSPIRE as of 01 Mar 2017


%\cite{Kagan:1989fp}
\bibitem{Kagan:1989fp} 
  A.~L.~Kagan,
  %``Radiative Quark Mass and Mixing Hierarchies From Supersymmetric Models With a Fourth Mirror Family,''
  Phys.\ Rev.\ D {\bf 40}, 173 (1989).
 % doi:10.1103/PhysRevD.40.173
  %%CITATION = doi:10.1103/PhysRevD.40.173;%%
  %26 citations counted in INSPIRE as of 01 Mar 2017

%\cite{Baumgart:2014jya}
\bibitem{Baumgart:2014jya} 
  M.~Baumgart, D.~Stolarski and T.~Zorawski,
  %``Split supersymmetry radiates flavor,''
  Phys.\ Rev.\ D {\bf 90}, no. 5, 055001 (2014)
%  doi:10.1103/PhysRevD.90.055001
  [arXiv:1403.6118 [hep-ph]].
  %%CITATION = doi:10.1103/PhysRevD.90.055001;%%
  %14 citations counted in INSPIRE as of 17 Feb 2017
%\cite{Altmannshofer:2014qha}\cite{Ma:2014yka}
\bibitem{Altmannshofer:2014qha} 
  W.~Altmannshofer, C.~Frugiuele and R.~Harnik,
  %``Fermion Hierarchy from Sfermion Anarchy,''
  JHEP {\bf 1412}, 180 (2014)
%  doi:10.1007/JHEP12(2014)180
  [arXiv:1409.2522 [hep-ph]].

%%%%%%%%%%%%%%%%%%%%%%%%%%%%%%%%%%%%%%%%
%%%%%%Added papers%%%%%%%%%%%%%%%%%%%%%%%%
%%%%%%%%%%%%%%%%%%%%%%%%%%%%%%%%%%%%%%%
%\cite{Borzumati:1999sp,Ferrandis:2004ng,Ferrandis:2004ri,Crivellin:2010ty,Crivellin:2011sj,Thalapillil:2014kya}
\bibitem{Borzumati:1999sp} 
  F.~Borzumati, G.~R.~Farrar, N.~Polonsky and S.~D.~Thomas,
  %``Soft Yukawa couplings in supersymmetric theories,''
  Nucl.\ Phys.\ B {\bf 555}, 53 (1999)
%  doi:10.1016/S0550-3213(99)00328-4
  [hep-ph/9902443].
  %%CITATION = doi:10.1016/S0550-3213(99)00328-4;%%
  %163 citations counted in INSPIRE as of 18 Jul 2017

%\cite{Ferrandis:2004ng,Ferrandis:2004ri,Crivellin:2010ty,Crivellin:2011sj,Thalapillil:2014kya}
\bibitem{Ferrandis:2004ng} 
  J.~Ferrandis,
  %``Radiative mass generation and suppression of supersymmetric contributions to flavor changing processes,''
  Phys.\ Rev.\ D {\bf 70}, 055002 (2004)
 % doi:10.1103/PhysRevD.70.055002
  [hep-ph/0404068].
  %%CITATION = doi:10.1103/PhysRevD.70.055002;%%
  %12 citations counted in INSPIRE as of 18 Jul 2017
  
  %\cite{Ferrandis:2004ri,Crivellin:2010ty,Crivellin:2011sj,Thalapillil:2014kya}
\bibitem{Ferrandis:2004ri} 
  J.~Ferrandis and N.~Haba,
  %``Supersymmetry breaking as the origin of flavor,''
  Phys.\ Rev.\ D {\bf 70}, 055003 (2004)
%  doi:10.1103/PhysRevD.70.055003
  [hep-ph/0404077].
  %%CITATION = doi:10.1103/PhysRevD.70.055003;%%
  %27 citations counted in INSPIRE as of 18 Jul 2017
  
  %\cite{Crivellin:2010ty,Crivellin:2011sj,Thalapillil:2014kya}
\bibitem{Crivellin:2010ty} 
  A.~Crivellin, J.~Girrbach and U.~Nierste,
  %``Yukawa coupling and anomalous magnetic moment of the muon: an update for the LHC era,''
  Phys.\ Rev.\ D {\bf 83}, 055009 (2011)
%  doi:10.1103/PhysRevD.83.055009
  [arXiv:1010.4485 [hep-ph]].
  %%CITATION = doi:10.1103/PhysRevD.83.055009;%%
  %24 citations counted in INSPIRE as of 18 Jul 2017
  %\cite{Crivellin:2011sj,Thalapillil:2014kya}
\bibitem{Crivellin:2011sj} 
  A.~Crivellin, L.~Hofer, U.~Nierste and D.~Scherer,
  %``Phenomenological consequences of radiative flavor violation in the MSSM,''
  Phys.\ Rev.\ D {\bf 84}, 035030 (2011)
 % doi:10.1103/PhysRevD.84.035030
  [arXiv:1105.2818 [hep-ph]].
  %%CITATION = doi:10.1103/PhysRevD.84.035030;%%
  %42 citations counted in INSPIRE as of 18 Jul 2017
  
  
  %\cite{Thalapillil:2014kya}
\bibitem{Thalapillil:2014kya} 
  A.~Thalapillil and S.~Thomas,
  %``Higgs Boson Yukawa Form Factors from Supersymmetric Radiative Fermion Masses,''
  arXiv:1411.7362 [hep-ph].
  %%CITATION = ARXIV:1411.7362;%%
  %2 citations counted in INSPIRE as of 18 Jul 2017
  
 %%%%%%%%%%%%flavor symmetric models%%%%%%%%%%%%%%%%% 


\bibitem{He:1989er} 
  X.~G.~He, R.~R.~Volkas and D.~D.~Wu,
  %``Radiative Generation of Quark and Lepton Mass Hierarchies From a Top Quark Mass Seed,''
  Phys.\ Rev.\ D {\bf 41}, 1630 (1990).
  %doi:10.1103/PhysRevD.41.1630
  %%CITATION = doi:10.1103/PhysRevD.41.1630;%%
  %29 citations counted in INSPIRE as of 17 Feb 2017
%\cite{Ma:2013mga}


\bibitem{Ma:2013mga} 
  E.~Ma,
  %``Radiative Origin of All Quark and Lepton Masses through Dark Matter with Flavor Symmetry,''
  Phys.\ Rev.\ Lett.\  {\bf 112}, 091801 (2014)
%  doi:10.1103/PhysRevLett.112.091801
  [arXiv:1311.3213 [hep-ph]].
  %%CITATION = doi:10.1103/PhysRevLett.112.091801;%%
  %47 citations counted in INSPIRE as of 17 Feb 2017






%\cite{Ma:2014yka}
\bibitem{Ma:2014yka} 
  E.~Ma,
  %``Syndetic Model of Fundamental Interactions,''
  Phys.\ Lett.\ B {\bf 741}, 202 (2015)
%  doi:10.1016/j.physletb.2014.12.045
  [arXiv:1411.6679 [hep-ph]].
  %%CITATION = doi:10.1016/j.physletb.2014.12.045;%%
  %17 citations counted in INSPIRE as of 14 Feb 2017

  %%CITATION = doi:10.1007/JHEP12(2014)180;%%
  %12 citations counted in INSPIRE as of 14 Feb 2017
%\cite{Ibarra:2014pfa}\cite{Altmannshofer:2014qha}\cite{Ma:2014yka}
%\bibitem{Ibarra:2014pfa} 
 % A.~Ibarra and A.~Solaguren-Beascoa,
  %``Lepton parameters in the see-saw model extended by one extra Higgs doublet,''
 % JHEP {\bf 1411}, 089 (2014)
%  doi:10.1007/JHEP11(2014)089
 % [arXiv:1409.5011 [hep-ph]].
  %%CITATION = doi:10.1007/JHEP11(2014)089;%%
  %5 citations counted in INSPIRE as of 14 Feb 2017
%\cite{Nomura:2016emz}\cite{Ibarra:2014pfa}\cite{Altmannshofer:2014qha}\cite{Ma:2014yka}
\bibitem{Nomura:2016emz} 
  T.~Nomura and H.~Okada,
  %``Radiatively induced Quark and Lepton Mass Model,''
  Phys.\ Lett.\ B {\bf 761}, 190 (2016)
 % doi:10.1016/j.physletb.2016.08.023
  [arXiv:1606.09055 [hep-ph]].
  %%CITATION = doi:10.1016/j.physletb.2016.08.023;%%
  %9 citations counted in INSPIRE as of 14 Feb 2017
%\cite{Natale:2016xob}\cite{Nomura:2016emz}\cite{Ibarra:2014pfa}\cite{Altmannshofer:2014qha}\cite{Ma:2014yka}
\bibitem{Natale:2016xob} 
  A.~Natale,
  %``A Radiative Model of Quark Masses with Binary Tetrahedral Symmetry,''
  Nucl.\ Phys.\ B {\bf 914}, 201 (2017)
%  doi:10.1016/j.nuclphysb.2016.11.006
  [arXiv:1608.06999 [hep-ph]].
  %%CITATION = doi:10.1016/j.nuclphysb.2016.11.006;%%
%\cite{CarcamoHernandez:2016pdu}
%\cite{Kownacki:2016hpm}
\bibitem{Kownacki:2016hpm} 
  C.~Kownacki and E.~Ma,
  %``Gauge $U(1)$ dark symmetry and radiative light fermion masses,''
  Phys.\ Lett.\ B {\bf 760}, 59 (2016)
  %doi:10.1016/j.physletb.2016.06.024
  [arXiv:1604.01148 [hep-ph]].
  %%CITATION = doi:10.1016/j.physletb.2016.06.024;%%
  %12 citations counted in INSPIRE as of 17 Feb 2017

\bibitem{CarcamoHernandez:2016pdu} 
  A.~E.~C\'arcamo Hern\'andez, S.~Kovalenko and I.~Schmidt,
  %``Radiatively generated hierarchy of lepton and quark masses,''
  arXiv:1611.09797 [hep-ph].
  %%CITATION = ARXIV:1611.09797;%%

%%%%%%%%%%%%%%%%%%%%%%%%%%%%%%%%%%

%\cite{Aprile:2015uzo}
\bibitem{XENON1T} 
  E.~Aprile {\it et al.} [XENON Collaboration],
  %``Physics reach of the XENON1T dark matter experiment,''
  JCAP {\bf 1604}, no. 04, 027 (2016)
%  doi:10.1088/1475-7516/2016/04/027
  [arXiv:1512.07501 [physics.ins-det]].
  %%CITATION = doi:10.1088/1475-7516/2016/04/027;%%
  %149 citations counted in INSPIRE as of 16 Mar 2017
%%%%%%%%%%%%%%%%%%%%%%%%%%%%%%%%%%%
\bibitem{PDG}
  C.~Patrignani {\it et al.} [Particle Data Group],
  %``Review of Particle Physics,''
  Chin.\ Phys.\ C {\bf 40}, no. 10, 100001 (2016).
%  doi:10.1088/1674-1137/40/10/100001
  %%CITATION = doi:10.1088/1674-1137/40/10/100001;%%
  %403 citations counted in INSPIRE as of 24 Feb 2017

%\bibitem{PDG}
%  K.~A.~Olive \textit{et al.} [Paricle Data Group],
%  Chin.\ Phys.\ C\ {\bf 38}, 090001 (2014).

\bibitem{CKMfitter}
 CKMfitter global fit results as of Summer 2016 (ICHEP 2016 conference)\\
 \texttt{http://ckmfitter.in2p3.fr/www/html/ckm\_main.html}


\bibitem{RunDec}
  K.~G.~Chetyrkin, J.~H.~Kuhn and M.~Steinhauser,
  %``RunDec: A Mathematica package for running and decoupling of the strong coupling and quark masses,''
  Comput.\ Phys.\ Commun.\  {\bf 133}, 43 (2000)
  %doi:10.1016/S0010-4655(00)00155-7
  [hep-ph/0004189].
  %%CITATION = doi:10.1016/S0010-4655(00)00155-7;%%
  %245 citations counted in INSPIRE as of 01 f\UTF{00E9}vr. 2016

\bibitem{SMrun}
  H.~Arason, D.~J.~Castano, B.~Keszthelyi, S.~Mikaelian, E.~J.~Piard, P.~Ramond and B.~D.~Wright,
  %``Renormalization group study of the standard model and its extensions. 1. The Standard model,''
  Phys.\ Rev.\ D {\bf 46}, 3945 (1992).
  %doi:10.1103/PhysRevD.46.3945
  %%CITATION = doi:10.1103/PhysRevD.46.3945;%%
  %357 citations counted in INSPIRE as of 01 f\UTF{00E9}vr. 2016


%\cite{Abe:2016wck}
\bibitem{Okawa} 
  T.~Abe, J.~Kawamura, S.~Okawa and Y.~Omura,
  %``Dark matter physics, flavor physics and LHC constraints in the dark matter model with a bottom partner,''
  JHEP {\bf 1703}, 058 (2017)
%  doi:10.1007/JHEP03(2017)058
  [arXiv:1612.01643 [hep-ph]].
  %%CITATION = doi:10.1007/JHEP03(2017)058;%%
  %1 citations counted in INSPIRE as of 23 Mar 2017

%\bibitem{CKMf15}
% Web site of CKMfitter group (EPS-HEP 2015 conference)\\
 %\texttt{http://ckmfitter.in2p3.fr/www/results/plots\_eps15/num/ckmEval\_ results\_eps15.html}


%%\cite{Buras:2012jb}\cite{Buras:2004ub}\cite{Gorbahn:2006bm}
%\bibitem{Buras:2012jb} 
%  A.~J.~Buras, F.~De Fazio and J.~Girrbach,
%  %``The Anatomy of Z' and Z with Flavour Changing Neutral Currents in the Flavour Precision Era,''
%  JHEP {\bf 1302}, 116 (2013)
%%  doi:10.1007/JHEP02(2013)116
%  [arXiv:1211.1896 [hep-ph]].
%  %%CITATION = doi:10.1007/JHEP02(2013)116;%%
%  %70 citations counted in INSPIRE as of 22 Dec 2015

\bibitem{Lattice1}
  Jack~Laiho, E.~Lunghi and Ruth~S.~Van~de~Water,
  %``Lattice QCD inputs to the CKM unitarity triangle analysis,''
  Phys.\ Rev.\ D {\bf 81}, 034503 (2010) [arXiv:0910.2928 [hep-ph]].
  %%CITATION = ARXIV:0910.2928;%%
  %249 citations counted in INSPIRE as of 30 Nov 2015
  
%  See also the latest values in \verb|http://www.latticeaverages.org|.

%\cite{Aoki:2016frl}
\bibitem{Lattice2} 
  S.~Aoki {\it et al.},
  %``Review of lattice results concerning low-energy particle physics,''
  Eur.\ Phys.\ J.\ C {\bf 77}, no. 2, 112 (2017)
%  doi:10.1140/epjc/s10052-016-4509-7
  [arXiv:1607.00299 [hep-lat]].
  %%CITATION = doi:10.1140/epjc/s10052-016-4509-7;%%
  %92 citations counted in INSPIRE as of 01 Mar 2017



%\cite{Brod:2011ty}\cite{Buras:1990fn}\cite{Brod:2010mj}
\bibitem{Brod:2011ty} 
  J.~Brod and M.~Gorbahn,
  %``Next-to-Next-to-Leading-Order Charm-Quark Contribution to the CP Violation Parameter epsilon_K and Delta M_K,''
  Phys.\ Rev.\ Lett.\  {\bf 108}, 121801 (2012)
%  doi:10.1103/PhysRevLett.108.121801
  [arXiv:1108.2036 [hep-ph]].
  %%CITATION = doi:10.1103/PhysRevLett.108.121801;%%
  %89 citations counted in INSPIRE as of 25 Dec 2015

%\cite{Buras:1990fn}\cite{Brod:2010mj}
\bibitem{Buras:1990fn} 
  A.~J.~Buras, M.~Jamin and P.~H.~Weisz,
  %``Leading and Next-to-leading {QCD} Corrections to $\epsilon$ Parameter and $B^0 - \bar{B}^0$ Mixing in the Presence of a Heavy Top Quark,''
  Nucl.\ Phys.\ B {\bf 347}, 491 (1990).
 % doi:10.1016/0550-3213(90)90373-L
  %%CITATION = doi:10.1016/0550-3213(90)90373-L;%%
  %616 citations counted in INSPIRE as of 25 Dec 2015

%\cite{Brod:2010mj}
\bibitem{Brod:2010mj} 
  J.~Brod and M.~Gorbahn,
  %``Epsilon_K at Next-to-Next-to-Leading Order: The Charm-Top-Quark Contribution,''
  Phys.\ Rev.\ D {\bf 82}, 094026 (2010)
%  doi:10.1103/PhysRevD.82.094026
  [arXiv:1007.0684 [hep-ph]].
  %%CITATION = doi:10.1103/PhysRevD.82.094026;%%
  %98 citations counted in INSPIRE as of 25 Dec 2015



%\cite{Buchalla:1998ba}
\bibitem{Buchalla:1998ba} 
  G.~Buchalla and A.~J.~Buras,
  %``The rare decays $K\to \pi \nu\bar\nu$, $B \to X \nu\bar\nu$ and $B \to l^+ l^-$: An Update,''
  Nucl.\ Phys.\ B {\bf 548}, 309 (1999)
 % doi:10.1016/S0550-3213(99)00149-2
  [hep-ph/9901288].
  %%CITATION = doi:10.1016/S0550-3213(99)00149-2;%%
  %345 citations counted in INSPIRE as of 20 Jun 2016


  %\cite{Lubicz:2008am}
\bibitem{Lubicz:2008am} 
  V.~Lubicz and C.~Tarantino,
  %``Flavour physics and Lattice QCD: Averages of lattice inputs for the Unitarity Triangle Analysis,''
  Nuovo Cim.\ B {\bf 123}, 674 (2008)
 % doi:10.1393/ncb/i2008-10650-3
  [arXiv:0807.4605 [hep-lat]].
  %%CITATION = doi:10.1393/ncb/i2008-10650-3;%%
  %86 citations counted in INSPIRE as of 16 Feb 2017
  %\cite{Hisano:2015pma}{Hisano:2016afc} 

%%%%%%%%%%%%%%
%%%%%%EDM%%%%%%%%%%%%


%\cite{Afach:2015sja}
\bibitem{EDM} 
  J.~M.~Pendlebury {\it et al.},
  %``Revised experimental upper limit on the electric dipole moment of the neutron,''
  Phys.\ Rev.\ D {\bf 92}, no. 9, 092003 (2015)
 % doi:10.1103/PhysRevD.92.092003
  [arXiv:1509.04411 [hep-ex]].
  %%CITATION = doi:10.1103/PhysRevD.92.092003;%%
  %50 citations counted in INSPIRE as of 06 Mar 2017


\bibitem{Amhis:2014hma} 
  Y.~Amhis {\it et al.} [Heavy Flavor Averaging Group (HFAG)],
  %``Averages of $b$-hadron, $c$-hadron, and $\tau$-lepton properties as of summer 2014,''
  arXiv:1412.7515 [hep-ex].
  %%CITATION = ARXIV:1412.7515;%%
  %378 citations counted in INSPIRE as of 30 Dec 2016      \\ 


%\cite{Hisano:2012sc}
\bibitem{Hisano:2012sc} 
  J.~Hisano, J.~Y.~Lee, N.~Nagata and Y.~Shimizu,
  %``Reevaluation of Neutron Electric Dipole Moment with QCD Sum Rules,''
  Phys.\ Rev.\ D {\bf 85}, 114044 (2012)
 % doi:10.1103/PhysRevD.85.114044
  [arXiv:1204.2653 [hep-ph]].
  %%CITATION = doi:10.1103/PhysRevD.85.114044;%%
  %42 citations counted in INSPIRE as of 06 Mar 2017
\bibitem{Fuyuto:2012yf} 
  K.~Fuyuto, J.~Hisano and N.~Nagata,
  %``Neutron electric dipole moment induced by strangeness revisited,''
  Phys.\ Rev.\ D {\bf 87}, no. 5, 054018 (2013)
 % doi:10.1103/PhysRevD.87.054018
  [arXiv:1211.5228 [hep-ph]].
  %%CITATION = doi:10.1103/PhysRevD.87.054018;%%
  %25 citations counted in INSPIRE as of 06 Mar 2017
  %\cite{Jung:2013hka}
\bibitem{Jung:2013hka} 
  M.~Jung and A.~Pich,
  %``Electric Dipole Moments in Two-Higgs-Doublet Models,''
  JHEP {\bf 1404}, 076 (2014)
 % doi:10.1007/JHEP04(2014)076
  [arXiv:1308.6283 [hep-ph]].
  %%CITATION = doi:10.1007/JHEP04(2014)076;%%
  %63 citations counted in INSPIRE as of 06 Mar 2017

\bibitem{permanentEDM} 
B. Graner, Y. Chen, E. G. Lindahl, B. R. Heckel,
Phys. Rev. Lett. 116, 161601 (2016)
[arXiv:1601.04339 [Atomic Physics]].

%%%%%%%%%%%%%%
%%%%%%DM%%%%%%%%%%%%



%\cite{Escudero:2016gzx}
\bibitem{Higgsportal} 
  M.~Escudero, A.~Berlin, D.~Hooper and M.~X.~Lin,
  %``Toward (Finally!) Ruling Out Z and Higgs Mediated Dark Matter Models,''
  JCAP {\bf 1612}, 029 (2016)
%  doi:10.1088/1475-7516/2016/12/029
  [arXiv:1609.09079 [hep-ph]].
  %%CITATION = doi:10.1088/1475-7516/2016/12/029;%%
  %9 citations counted in INSPIRE as of 16 Mar 2017


%\cite{Griest:1989wd}
\bibitem{Griest:1989wd} 
  K.~Griest and M.~Kamionkowski,
  %``Unitarity Limits on the Mass and Radius of Dark Matter Particles,''
  Phys.\ Rev.\ Lett.\  {\bf 64}, 615 (1990).
%  doi:10.1103/PhysRevLett.64.615
  %%CITATION = doi:10.1103/PhysRevLett.64.615;%%
  %383 citations counted in INSPIRE as of 16 Mar 2017

%\cite{Akerib:2015rjg}
\bibitem{LUX2015} 
  D.~S.~Akerib {\it et al.} [LUX Collaboration],
  %``Improved Limits on Scattering of Weakly Interacting Massive Particles from Reanalysis of 2013 LUX Data,''
  Phys.\ Rev.\ Lett.\  {\bf 116}, no. 16, 161301 (2016)
%  doi:10.1103/PhysRevLett.116.161301
  [arXiv:1512.03506 [astro-ph.CO]].
  %%CITATION = doi:10.1103/PhysRevLett.116.161301;%%
  %197 citations counted in INSPIRE as of 02 Nov 2016

%\cite{Akerib:2016vxi}
\bibitem{LUX2016} 
  D.~S.~Akerib {\it et al.},
  %``Results from a search for dark matter in the complete LUX exposure,''
  arXiv:1608.07648 [astro-ph.CO].
  %%CITATION = ARXIV:1608.07648;%%
  %51 citations counted in INSPIRE as of 02 Nov 2016

%\cite{Tan:2016zwf}
\bibitem{Panda} 
  A.~Tan {\it et al.} [PandaX-II Collaboration],
  %``Dark Matter Results from First 98.7 Days of Data from the PandaX-II Experiment,''
  Phys.\ Rev.\ Lett.\  {\bf 117}, no. 12, 121303 (2016)
%  doi:10.1103/PhysRevLett.117.121303
  [arXiv:1607.07400 [hep-ex]].
  %%CITATION = doi:10.1103/PhysRevLett.117.121303;%%
  %47 citations counted in INSPIRE as of 02 Nov 2016



%\cite{Arhrib:2013ela}
\bibitem{IDM0} 
  A.~Arhrib, Y.~L.~S.~Tsai, Q.~Yuan and T.~C.~Yuan,
  %``An Updated Analysis of Inert Higgs Doublet Model in light of the Recent Results from LUX, PLANCK, AMS-02 and LHC,''
  JCAP {\bf 1406}, 030 (2014)
%  doi:10.1088/1475-7516/2014/06/030
  [arXiv:1310.0358 [hep-ph]].
  %%CITATION = doi:10.1088/1475-7516/2014/06/030;%%
  %77 citations counted in INSPIRE as of 16 Mar 2017
%\cite{Diaz:2015pyv}


%\cite{Goudelis:2013uca}
\bibitem{IDM0-1} 
  A.~Goudelis, B.~Herrmann and O.~Stal,
  %``Dark matter in the Inert Doublet Model after the discovery of a Higgs-like boson at the LHC,''
  JHEP {\bf 1309}, 106 (2013)
%  doi:10.1007/JHEP09(2013)106
  [arXiv:1303.3010 [hep-ph]].
  %%CITATION = doi:10.1007/JHEP09(2013)106;%%
  %102 citations counted in INSPIRE as of 16 Mar 2017
  %\cite{Ilnicka:2015jba}
\bibitem{IDM0-2} 
  A.~Ilnicka, M.~Krawczyk and T.~Robens,
  %``Inert Doublet Model in light of LHC Run I and astrophysical data,''
  Phys.\ Rev.\ D {\bf 93}, no. 5, 055026 (2016)
%  doi:10.1103/PhysRevD.93.055026
  [arXiv:1508.01671 [hep-ph]].
  %%CITATION = doi:10.1103/PhysRevD.93.055026;%%
  %32 citations counted in INSPIRE as of 16 Mar 2017
  \bibitem{IDM0-3} 
  M.~A.~D\'iaz, B.~Koch and S.~Urrutia-Quiroga,
  %``Constraints to Dark Matter from Inert Higgs Doublet Model,''
  Adv.\ High Energy Phys.\  {\bf 2016}, 8278375 (2016)
%  doi:10.1155/2016/8278375
  [arXiv:1511.04429 [hep-ph]].
  %%CITATION = doi:10.1155/2016/8278375;%%
  %16 citations counted in INSPIRE as of 16 Mar 2017
%\cite{Belyaev:2016lok}
  
\bibitem{IDM} 
  A.~Belyaev, G.~Cacciapaglia, I.~P.~Ivanov, F.~Rojas and M.~Thomas,
  %``Anatomy of the Inert Two Higgs Doublet Model in the light of the LHC and non-LHC Dark Matter Searches,''
  arXiv:1612.00511 [hep-ph].
  %%CITATION = ARXIV:1612.00511;%%
  %4 citations counted in INSPIRE as of 16 Mar 2017
%%%%%%%%%%%%%%%%%%%%%%%%%%




%\cite{Charles:2013aka}
%\bibitem{Charles:2013aka} 
%  J.~Charles, S.~Descotes-Genon, Z.~Ligeti, S.~Monteil, M.~Papucci and K.~Trabelsi,
  %``Future sensitivity to new physics in $B_d, B_s$, and K mixings,''
%  Phys.\ Rev.\ D {\bf 89}, no. 3, 033016 (2014)
%  [arXiv:1309.2293 [hep-ph]].
  %%CITATION = doi:10.1103/PhysRevD.89.033016;%%
  %34 citations counted in INSPIRE as of 10 Dec 2015



%\cite{Bazavov:2016nty}
%\bibitem{Bazavov:2016nty} 
 % A.~Bazavov {\it et al.} [Fermilab Lattice and MILC Collaborations],
  %``$B^0_{(s)}$-mixing matrix elements from lattice QCD for the Standard Model and beyond,''
%  Phys.\ Rev.\ D {\bf 93}, no. 11, 113016 (2016)
%  doi:10.1103/PhysRevD.93.113016
%  [arXiv:1602.03560 [hep-lat]].
  %%CITATION = doi:10.1103/PhysRevD.93.113016;%%
  %18 citations counted in INSPIRE as of 14 Jul 2016


\bibitem{Okawa2}
Work in preparation.


\bibitem{Hisano:2015pma} 
  J.~Hisano, Y.~Muramatsu, Y.~Omura and M.~Yamanaka,
  %``Flavor violating $Z'$ from $SO(10)$ SUSY GUT in High-Scale SUSY,''
  Phys.\ Lett.\ B {\bf 744}, 395 (2015)
%  doi:10.1016/j.physletb.2015.04.020
  [arXiv:1503.06156 [hep-ph]].
  %%CITATION = doi:10.1016/j.physletb.2015.04.020;%%
  %4 citations counted in INSPIRE as of 18 Feb 2017


\bibitem{Hisano:2016afc} 
  J.~Hisano, Y.~Muramatsu, Y.~Omura and Y.~Shigekami,
  %``Flavor physics induced by light $Z'$ from SO(10) GUT,''
  JHEP {\bf 1611}, 018 (2016)
%  doi:10.1007/JHEP11(2016)018
  [arXiv:1607.05437 [hep-ph]].
  %%CITATION = doi:10.1007/JHEP11(2016)018;%%
  %\cite{Hisano:2015pma}

\end{thebibliography}
\end{document}